\documentclass[letter,12pt]{article}

\usepackage{amsmath}
\usepackage{amssymb}
\usepackage[nodisplayskipstretch]{setspace} 
\usepackage[top=1in, bottom=1in, left=1in, right=1in]{geometry}
\usepackage{graphicx}
\usepackage{placeins}
\usepackage[labelfont=bf,labelsep=period,font=footnotesize,width=6.5in]{caption}
\usepackage{lineno}
\usepackage{natbib}
\usepackage{longtable}
\usepackage{color}
\usepackage{caption}
\usepackage{subcaption}
\newenvironment{myenumerate}
{ \begin{enumerate}
    \setlength{\itemsep}{0pt}
    \setlength{\parskip}{0pt}
    \setlength{\parsep}{0pt}     }
{ \end{enumerate}                  } 

\newcommand{\beginsupplement}{%
        \setcounter{table}{0}
        \renewcommand{\thetable}{S\arabic{table}}%
        \setcounter{figure}{0}
        \renewcommand{\thefigure}{S\arabic{figure}}%
     }

\bibpunct{(}{)}{,}{a}{}{,}

\raggedright
\makeatletter
\renewcommand\section{\@startsection{section}{1}{0in}{0.1\baselineskip}{0.1\baselineskip}{\normalfont\normalsize\bfseries}} 
\makeatother

\makeatletter
\renewcommand\subsection{\@startsection{subsection}{1}{-\parindent}{0.1\baselineskip}{0.1\baselineskip}{\normalfont\normalsize\textit}}
\makeatother

\begin{document}

\begin{centering}
\ \\

\vspace{1.5in}
\Large{Balls, cups, and quasi-potentials: quantifying stability in stochastic systems}\\
\vspace{1in}
\normalsize{Ben C.~Nolting and Karen C.~Abbott }\\ \vspace{0.2in}
Department of Biology\\
Case Western Reserve University\\
Cleveland, OH 44106 U.S.A\\  \vspace{0.2in}
\end{centering}
\vspace{\fill}
\noindent \today \ draft\\
\noindent Manuscript accepted at Ecology
\newpage

\section*{Abstract}

When a system has more than one stable state, how can the stability of these states be compared? This deceptively simple question has important consequences for ecosystems, because systems with alternative stable states can undergo dramatic regime shifts. The probability, frequency, duration, and dynamics of these shifts will all depend on the relative stability of the stable states. Unfortunately, the concept of ``stability" in ecology has suffered from substantial confusion and this is particularly problematic for systems where stochastic perturbations can cause shifts between coexisting alternative stable states. A useful way to visualize stable states in stochastic systems is with a ball\-in\-cup diagram, in which the state of the system is represented as the position of a ball rolling on a surface, and the random perturbations can push the ball from one basin of attraction to another. The surface is determined by a potential function, which provides a natural stability metric. However, systems amenable to this representation, called gradient systems, are quite rare. As a result, the potential function is not widely used and other approaches based on linear stability analysis have become standard.  Linear stability analysis is designed for local analysis of deterministic systems and, as we show, can produce a highly misleading picture of how the system will behave under continual, stochastic perturbations. In this paper, we show how the potential function can be generalized so that it can be applied broadly, employing a concept from stochastic analysis called the quasi\-potential. Using three classic ecological models, we demonstrate that the quasi\-potential provides a useful way to quantify stability in stochastic systems. We show that the quasi\-potential framework helps clarify long\-standing confusion about stability in stochastic ecological systems, and we argue that ecologists should adopt it as a practical tool for analyzing these systems.

\vspace{0.5in}
 \noindent{\it Keywords:} alternative stable states, stochastic dynamics, regime shifts, quasi\-potential, Freidlin\-Wentzell, stochastic differential equations, Hamilton\-Jacobi, resilience

\newpage
\section*{Introduction}
Researchers have long been fascinated by the possibility for ecosystems to have more than one stable state \citep{May:1977tm, Beisner:2003wu}. Such ecosystems have been observed in both natural \citep{vandeKoppel:2001wp} and experimental \citep{Chase:2003iv} settings. Systems with multiple (i.e., alternative) stable states can can abruptly shift from one stable state to another, sometimes with catastrophic consequences \citep{Scheffer:2003is}, so understanding their properties is crucially important.

Unfortunately, the understanding of alternative stable states has been significantly hampered by ambiguity about the term ``stable". \citet{Grimm:1997tg} note that stability is ``one of the most nebulous terms in the whole of ecology," and they catalog 163 different definitions. Much of this confusion arises when researchers attempt to apply tools designed for the analysis of deterministic models to stochastic models. Fortunately, there is a well\-developed mathematical framework, the Freidlin\-Wentzell quasi\-potential \citep{Freidlin:2012wd}, that provides a rigorous yet natural way to understand alternative stable states in stochastic systems. In this paper, we explain how this tool can clarify much of the confusion about stability in ecological systems by translating intuitive concepts into quantifiable mathematical properties.  Through three examples, we show how the quasi\-potential serves as a useful metric of stability, and allows for effective stability comparison between alternative stable states. The results from quasi\-potential analysis often contrast with those  from standard stability analysis, and our examples explore these discrepancies. Furthermore, the quasi\-potential allows for stability to be quantified on a continuum that corresponds well with the system's dynamics, and it can be applied to any system state, regardless of whether that state is a deterministic equilibrium. Using the quasi\-potential, a system can be decomposed into orthogonal components, and we explain how this decomposition can be interpreted ecologically. Finally, the quasi\-potential offers insight into the most probable paths a system will take in transitioning from one state to another.

Holling's foundational work on resilience and stability anticipated the quasi\-potential's basic essence \citep{Holling:1973wh}; later, \citet{Tuljapurkar:1979kd} made the insight that Holling's intuitive ideas were connected to the mathematical work of Freidlin and Wentzell \citeyearpar{Freidlin:1970wk}. At that time, numerical methods were insufficient to allow for general, practical computation of quasi\-potentials \citep*[see][]{Ludwig:1975tw}, so Tuljapurkar and Semura's insight did not receive the recognition it deserved. In subsequent decades, the flurry of research on alternative stable states largely overlooked this insight. Recently, the quasi\-potential has been embraced by researchers analyzing models in other areas of biology, although it often appears under other names, and is disconnected from the Freidlin\-Wentzell formulation (but see Zhou 2012). These applications include gene regulatory networks \citep{Lv:2014hp,Zhou:2012hz}, neural networks \citep{Yan:2013dx}, and evolution \citep{Zhang:2012ks, Wang:2011ef}. Very recently, it has been applied to a predator\-prey system \citep{Xu:2014hg}, and with countless other possibilities for application, we argue that the quasi\-potential is poised to become a major quantitative tool in ecology.

This paper makes three novel contributions to the field of ecology. First, it shows how the quasi\-potential can clarify the confusing tangle of stability concepts that confront ecologists. Second, it demonstrates how the quasi\-potential can be used to quantify stability in systems with alternative stable states, and how the results can be different from and often more useful than deterministic methods. Finally, it shows how a new numerical algorithm for the computation of quasi\-potentials \citep{Cameron:2012ex} can be expanded for application to systems with multiple stables states, and highlights the utility of the quasi\-potential for understanding such systems.

We use three well\-established ecological models to illustrate these ideas. First, we show how traditional linear stability analysis fails to capture the salient features of a stochastic lake eutrophication model, and explain how the system's potential function provides more useful analytic insights. Next, we move to higher\-dimensional systems, where potential functions rarely exist. We explore a consumer\-resource model with alternative stable states that does not have a potential function. We explain how the quasi\-potential is defined, and show its usefulness in analyzing this model. Finally, we explore another consumer\-resource model with a stable limit cycle to demonstrate how the quasi\-potential is useful  when stable states are more complicated than point equilibria. We conclude by discussing the quasi\-potential as a unifying framework for existing notions of stability in stochastic systems.
\medskip
\section*{Example 1: Lake Eutrophication}
Lake ecosystems are among the most well\-studied examples of alternative stable states in ecology. A foundational model by \citet{Carpenter:1999wt} successfully describes the coexistence of a eutrophic state, corresponding to high phosphorous concentration, and an oligotrophic state, corresponding to low phosphorous concentration. Later work by \citet{Guttal:2007dp} showed how stochasticity can cause this system to switch between the two stable states, and we will use their model as a starting point for exploring the quantification of stochastic stability.

The underlying deterministic model (i.e., the ``deterministic skeleton") describes how the nutrient (phosphorous) concentration $x$ changes over time:
\begin{equation}
\label{lakedet}
\frac{dx}{dt}=c-sx+r\frac{x^{q}}{x_{0}^{q}+x^{q}}.
\end{equation}
$c$ is the nutrient inflow rate and $s$ is the nutrient loss rate
(due to sedimentation, outflow, and sequestration
in benthic plants). The last term represents nutrient recycling. $r$ is the maximum recycling rate, $x_{0}$ is the half\-saturation constant, and $q$ specifies
the shape of the sigmoidal recycling curve. At
$s\!=\!1$, $r\!=\!1$, $x_{0}\!=\!1$, $q\!=\!8$, and $c\!=\!0.53$ (as in \citealt{Guttal:2007dp}),
the system has alternative stable states: a low phosphorous oligotrophic
state, $x_{L}\!=\!0.537$, and a high phosphorous eutrophic state,
$x_{H}\!=\!1.491$, separated by an unstable
equilibrium (a saddle), $x_{S}\!=\!0.971$.

The standard technique for studying systems like this one, is linear stability analysis. The eigenvalue of the linearized system at $x_{S}$ is $\lambda_{S}\!=\!1.032$, so it is an unstable equilibrium. The eigenvalues corresponding to $x_{L}$ and $x_{H}$ are $\lambda_{L}\!=\!-0.899$ and $\lambda_{H}\!=\!-0.797$, respectively, so both $x_{L}$ and $x_{H}$ are stable equilibria. The more negative the eigenvalue, the faster the return to the equilibrium following a small perturbation; $\lambda_{L}\!<\!\lambda_{H},$ so the linear analysis indicates that the oligotrophic state is more stable than the eutrophic state.
\medskip
\subsection*{Ball\-in\-cup}

An alternative approach to quantifying stability, and one that is fundamental to the theory of alternative stable
states, is the ``ball\-in\-cup'' heuristic \citep{Beisner:2003wu}.
In this framework, the state of the system is represented by the position
of a ball rolling on a surface. The ball rolls downhill, but is also
subject to continual, stochastically varying perturbations. In the absence
of perturbations, the ball will roll to the bottom
of a valley. Such locations correspond to stable equilibria
of the deterministic skeleton of the system ($x_L$ and $x_H$ in our example); a system with alternative
stable states has more than one valley. The ``cup'' is the area surrounding an equilibrium
that is attracted to it; this is called its domain (or basin) of attraction.

The ball\-in\-cup framework is not just a useful metaphor -- it can also
yield a mathematical description. For the lake system, define
\begin{equation} \label{Udef}
U(x)=-\int_{x_{H}}^{x}f(\xi)d\xi ,
\end{equation}
($\xi$ is a dummy variable for integration), so that the differential equation becomes: 
$
\frac{dx}{dt} = -U'(x)$.
The dynamics of this system turn out to be equivalent to a ball\-in\-cup
system with surface specified by the function $U$. In analogy with
the physics of the ball\-in\-cup metaphor, $U$ is called the ``potential
function" or simply the ``potential". For the lake system, this surface has local
minima at $x_{L}$ and $x_{H},$ as shown in figure~\ref{Fig1}a.

When random perturbations are present, the ball can
be jostled from one basin of attraction to another. Note that stochasticity
lies at the heart of the theory of alternative stable states. In a
purely deterministic system, the ball would roll to an equilibrium
and stay there. The presence or absence of other stable states
would be irrelevant, because the ball would have no way of visiting
them. Perhaps the surface could change over time, so that the basin of attraction
occupied by the ball ceases to be a basin, and the ball rolls out to a different stable state. This situation corresponds to a bifurcation
of the system's deterministic skeleton; the ball's transition requires the destruction of a stable state. In this paper, we are interested in
how systems can transition between \textit{coexisting} alternative stable states. Perturbations are required for the system to undergo these transitions; therefore, we argue that the appropriate framework for an alternative stable
state model is a stochastic one. Furthermore, real ecological systems are always subject to random perturbations. In order to apply the ball\-in\-cup heuristic to a perturbed system, we next demonstrate an approach to incorporating stochasticity into model~\eqref{lakedet}.
\medskip
\subsection*{Stochastic Differential Equation Model}

If the nutrient concentration varies randomly over time,
the lake can shift from one stable state to the other. To study this scenario, we translate the original deterministic model into a stochastic differential equation. A brief explanation of stochastic differential equation models is provided in appendix~\ref{subsec:SDEAppendix}, and more extensive accounts can be found in textbooks \citep[e.g.][]{Allen:2007ww}. Here, we give an informal description of the major concepts, and use discrete\-time analogies to avoid overly technical mathematical terminology.

To emphasize that nutrient concentration is now a stochastic process, and not just a deterministic function of time, we switch notation from $x(t)$ to $X\!(t)$. For each $t\!>\!0$, $x(t)$ is a number, but $X\!(t)$ is a random variable, which can take on any of a set of possible values according to probabilistic rules. A realization of the stochastic process is a deterministic function of time associated with a specific set of random events; this can be thought of as an observed time series, or the result of a single simulation run.

In the original model~\eqref{lakedet}, the external input of nutrients occurs at a constant rate $c$. In a small time interval $dt$, the external input is $c\, dt$. In reality, this input is likely to vary randomly; this is commonly modeled by adding a Gaussian white noise process, $dW\!(t)$ (``noise" is used synonymously with ``stochastic" or ``random"). At each $t\!>\!0$, $dW\!(t)$ is a normally distributed random variable with mean zero and variance $dt$. Since the values are independent of $t$, this is simply written as $dW$. The white noise process we describe here has no temporal autocorrelation, and its frequency spectrum is uniform -- the descriptor ``white" is used in analogy with white light. The accumulated change obtained by adding $dW$ over time yields a Wiener process, also known as Brownian motion. White noise is a useful starting point, but many applications require other types of noise; for example, colored noise might be used instead when perturbations are autocorrelated \citep[e.g.][]{Sharma:2014eh}. A discussion about generalizing the framework in this paper to different noise types is included in the Limitations and Generalizations section.

If the constant input rate $c$ is perturbed by a Gaussian white noise process with intensity $\sigma$ (analogous to the standard deviation in discrete time systems), then the external input in a small interval $dt$ is $c\, dt+\sigma dW$. The change in nutrient concentration over this time interval is given by
\begin{equation}
\label{lakestoch}
dX=\left(c-X+\frac{X^{q}}{1+X^{q}}\right)\,dt+\sigma\,dW.
\end{equation}
Again using equation \eqref{Udef} to define the potential, this system can equivalently be written as
\begin{equation}
\label{stochpot}
dX=-U'(X)\,dt+\sigma\,dW.
\end{equation}
In terms of the ball\-in\-cup heuristic, the shape
of the surface is specified by the potential function $U$, and this is
independent of $\sigma$. The noise intensity $\sigma$ only contributes to the movement of the ball on this surface, as determined by the last term in equation~\eqref{stochpot}.

We have described this model in terms of change over discrete time intervals, but it is also valid in the continuous time limit, $dt\!\rightarrow\!0$. For continuous time, which will be the focus of the rest of this paper,~\eqref{lakestoch} is called a stochastic differential equation. The notation in the stochastic differential equation $dX\!=\!\ldots$ is different than the deterministic differential equation notation $\frac{dx}{dt}=\ldots$, because the former must be defined using integral equations (the realizations of $W\!(t)$ are not differentiable anywhere, so $\frac{dW}{dt}$, and hence $\frac{dX}{dt}$, would not make sense. We use the It\^{o} integration scheme to define stochastic differential equations in this paper; see appendix~\ref{subsec:SDEAppendix}).
\medskip
\subsection*{Utility of the potential for understanding the stochastic lake eutrophication model}

\begin{figure}[ht]
\centering
\begin{subfigure}[b]{0.6\textwidth}
\caption{}
\includegraphics[width=\textwidth]{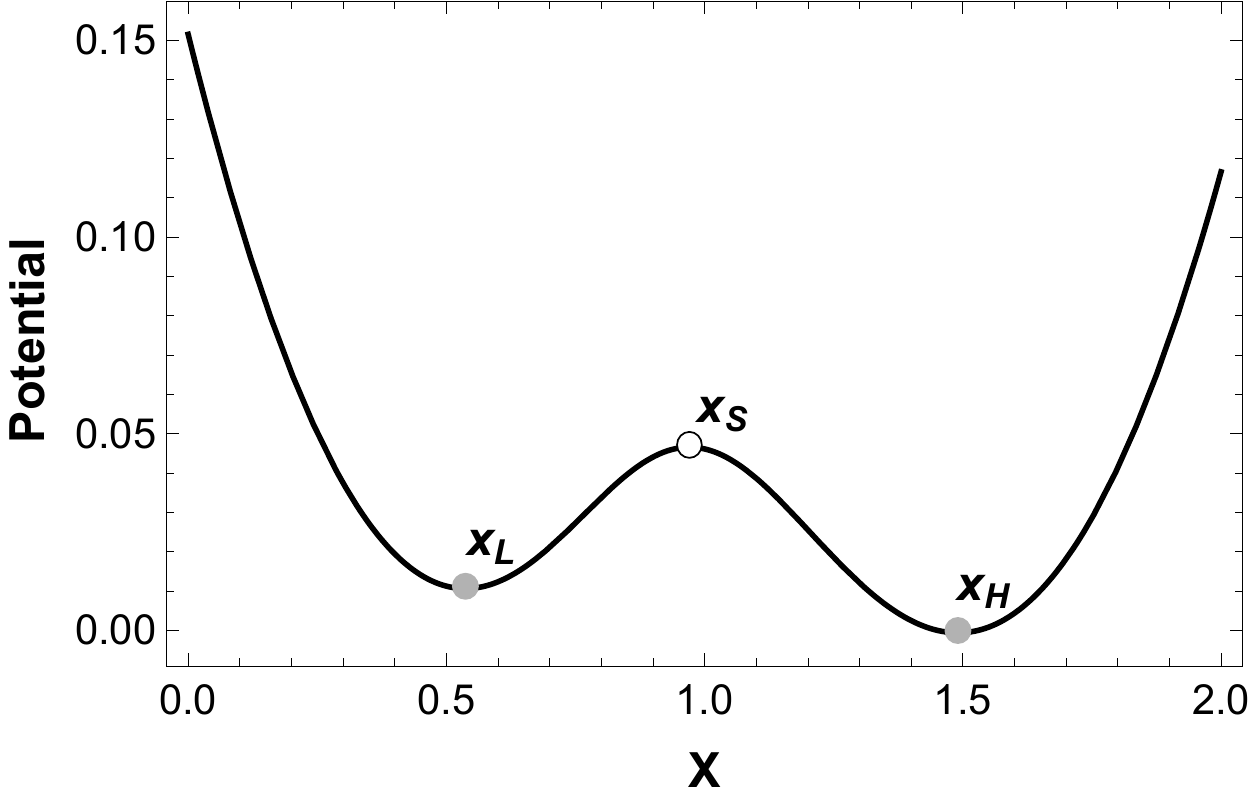}
\end{subfigure}
\vspace*{3mm}
\begin{subfigure}[b]{0.6\textwidth}
\caption{}
\includegraphics[width=\textwidth]{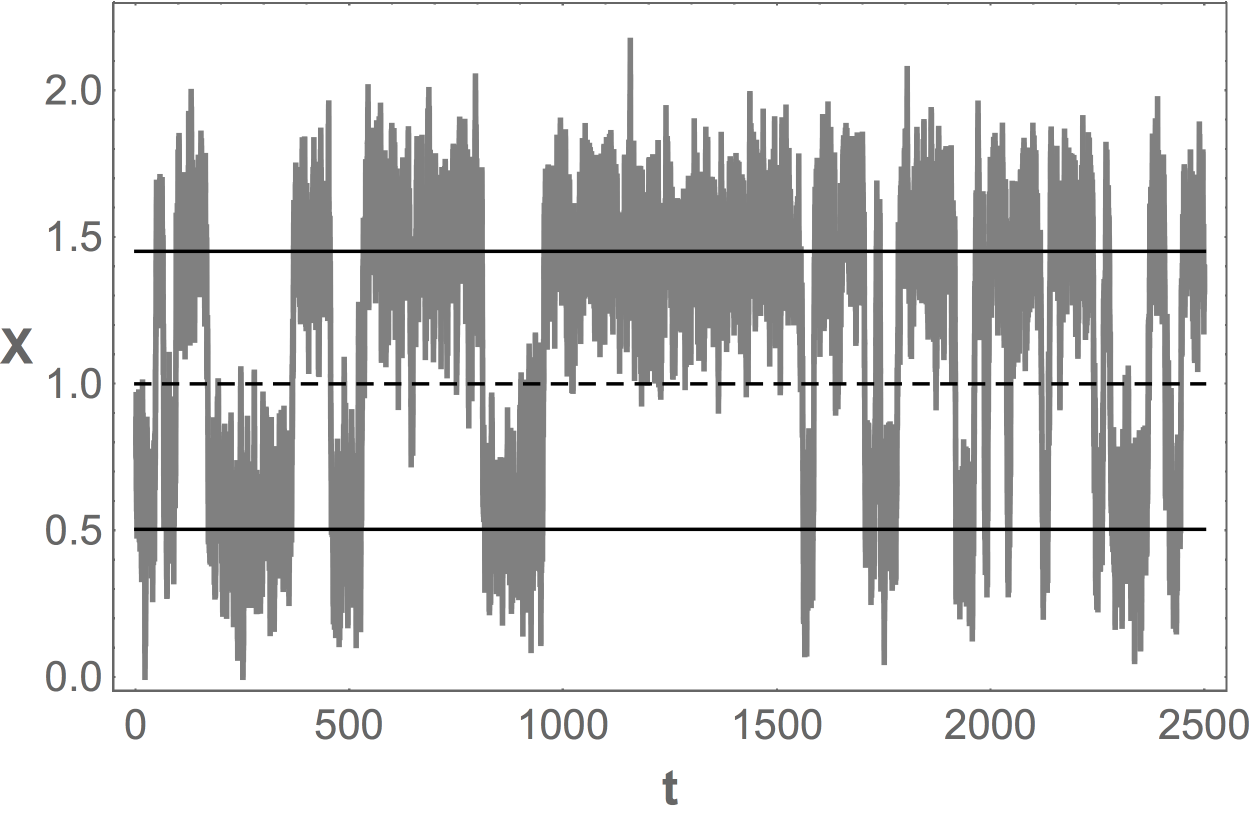}
\end{subfigure}
\caption{\small {\label{Fig1} Lake eutrophication model (example 1). \textbf{(a)} The potential function for equation~\eqref{lakedet}. The horizontal axis is the scaled nutrient (phosphorous) concentration and the vertical axis is the (dimensionless) potential. Gray disks are stable equilibria, and the white disk is an unstable (saddle) equilibrium. The dynamics of the system can be represented as a ball rolling on the surface specified by the potential function. Note that the basin around $x_{H}$ is deeper than that around $x_{L}$. \textbf{(b)} A realization of equation \eqref{lakestoch}, which models nutrient concentration, $x$, as a function of time, $t$. Variables are scaled, so the units are dimensionless. Integration was performed with the Euler-Maruyama method and $\Delta t=0.005$. The solid lines corresponds to stable equilibria $x_{L}$ (lower) and $x_{H}$ (higher) for the deterministic skeleton. The dashed line corresponds to the the saddle point $x_{S}$ of the deterministic skeleton. Note that the realization spends more time near $x_{H}$ than near $x_{L}$.}}
\end{figure}

One approach to understanding the stochastic lake eutrophication model is to calculate realizations (i.e. simulations) of~\eqref{lakestoch} for particular values of $\sigma$. This approach is limited, because it requires setting a particular $\sigma$; we will see later that the potential function provides a more general way of studying system dynamics. A realization with $\sigma=0.2$ is shown in figure~\ref{Fig1}b. All simulations in this paper were done with \textit{Mathematica}, and the code is available as a supplementary file. The realization in figure~\ref{Fig1}b, which is typical of realizations for this system with $\sigma=0.2$, 
switches between the two stable states. It spends more
time near $x_{H}$ than $x_{L}$; this suggests that the eutrophic (higher phosphorous) state
is more stable than the oligotrophic (lower phosphorous) state for this set of parameter
values. Note that this behavior is in contrast to the results of the linear
stability analysis of the deterministic skeleton. It is, however, in agreement with what the potential function tells us about the system, as we will demonstrate below.

For \eqref{lakestoch}, we find that $U\!(x_{L})\!=\!0.011$, $U\!(x_{S})\!=\!0.047$,
and $U\!(x_{H})\!=\!0$. Note that it is the relative, not the
absolute, values of the potential function that are important, so the
minimum value of the potential can be set at $0$. $U\!(x_{H})\!<\!U\!(x_{L})$,
so the potential function indicates that the eutrophic state is more
stable than the oligotrophic state. This corresponds to the intuitive
notion that we obtained from examining realizations like the one in figure~\ref{Fig1}b, but it contradicts
the results from the linear stability analysis. This discrepancy
arises because the linear stability analysis considers only
an infinitesimal neighborhood of an equilibrium. In the presence of
continuous stochastic perturbations, the system will leave such an
infinitesimal neighborhood, and the linear analysis of the skeleton breaks down. The linear analysis provides
information about the curvature of the potential surface at the bottom
of basins of attraction, but this information is purely local, in
that it does not take into account the larger geometry of the surface.
Therefore, the potential function provides a more appropriate measure
of stability for analyzing alternative stable states than linear stability analysis.

The potential function also relates to other important features of the stochastic system. The probability density function, $p(x,\!t)$, associated
with the random variable $X$ in~\eqref{lakestoch} describes the probability that $X\!(t)\!=\!x$. It is the solution to the Fokker\-Planck
equation: 
\begin{equation}
\frac{\partial p(x,t)}{\partial t}=\frac{\partial}{\partial x}\left(U'(x)\,p(x,t)\right)+\frac{\sigma^{2}}{2}\frac{\partial^{2}p(x,t)}{\partial x^{2}}.
\end{equation}
The steady\-state solution, $p_{s}(x)\!=\!\displaystyle{\lim_{t\rightarrow\infty}}p(x,\!t)$, is given by:
\begin{equation}
\label{steadystate}
p_{s}(x)=\frac{1}{Z}\exp\!\left(\!-\frac{2U(x)}{\sigma^{2}}\!\right),
\end{equation}
where $Z\!=\!\int_{0}^{\infty}\!\exp\left(-\frac{2\,U(x)}{\sigma^{2}}\right)\!dx$ is a normalization constant.
This equation shows that the steady\-state probability density is maximized
at the values of $x$ that minimize $U$, confirming that the minima (valleys) in $U$ correspond to the most likely system states.

The potential can be used
to gain insight about the time it takes the system to switch between
alternative stable states. If $\tau_{x_{L}}^{x_{H}}$ is the expected
time it takes a trajectory starting at $x_{L}$ to reach $x_{H}$,
(i.e., the mean first passage time), then \citep{kramers1940}: 
\begin{equation} \label{MFPT1}
\tau_{x_{L}}^{x_{H}}=\frac{2\pi}{\sqrt{U''(x_{L})\left|U''(x_{S})\right|}}\exp\left(\frac{2}{\sigma^{2}}\left(U(x_{S})-U(x_{L})\right)\right)\left(1+\mathcal{O}(\sigma)\right).
\end{equation}
Swapping $x_H$ for $x_L$ yields a comparable expression for the expected time to reach $x_{L}$ from $x_H$. The asymptotic notation $\mathcal{O}(\cdot)$ describes the error of the approximation as $\sigma\rightarrow 0$. The expected time for a trajectory to leave a basin of attraction around one of the stable states is thus largely dependent on the
depth of that basin -- the difference between peak $U$ (which occurs at the saddle equilibrium, $x_{S}$) and the value of $U$ at the stable equilibrium. 

The eigenvalue obtained in linear stability analysis describes the curvature of the potential at an equilibrium, equal to the second derivative of $U$; it determines the prefactor that multiplies the exponential function in equation~\eqref{MFPT1}. For a fixed valley depth, increased curvature is associated with decreased mean first passage time.  For instance, note that $\lambda_{L}\!=\!-U''\!(x_{L})$. As $x_{L}$ becomes more stable in the deterministic sense (i.e., as $\lambda_{L}$ becomes more negative), the curvature at $x_{L}$ increases, and the mean first passage time decreases (similar statements hold for $x_{H}$). At first glance, this seems counterintuitive  -- increasing stability is associated with decreased escape time -- but it makes sense because, for a fixed valley depth, increased curvature  decreases the horizontal distance between equilibria.

Knowledge about the
potential function thus provides information about the steady\-state
probability distribution, mean first passage times, and transition
frequencies, motivating its use as a
stability metric \citep{Zhou:2012hz, Wang:2011ef}. The potential function is especially useful because
it does not depend on the noise intensity $\sigma$ (in contrast to
the steady\-state probability distribution and mean first passage
times; see appendix~\ref{subsec:SmallNAppendix}).
\medskip
\section*{Example 2: Consumer and Resource With Alternative Stable States}

If the potential is so good at quantifying biologically\-relevant model behaviors, why isn't it routinely applied in ecology? Unfortunately, in most
cases, there will not exist a function $U$ that satisfies the mathematical definition of a potential (see appendix~\ref{subsec:FWAppendix}). Systems that have such a function are called ``gradient systems". One\-dimensional systems are always gradient systems, but systems with more than a single state variable almost never are. For non\-gradient systems, we cannot use a potential function
to quantify stability, as we did in the first example. It is for this reason that ecologists typically rely on approaches like linear stability analysis instead; although these approaches give more limited biological insights, they are more widely applicable mathematically. In what follows, we show how to generalize the potential for non\-gradient systems, thus allowing us to apply the many desirable features of potential analysis to a much broader range of ecological systems.

For an ecological example of a two\-dimensional non\-gradient system, we turn to a model of phytoplankton
and zooplankton populations. Let $R$ be the phytoplankton (resource) population density
and $C$ the zooplankton (consumer) population density. Using the deterministic skeleton of a standard plankton consumer\-resource model 
\citep{Collie:1994wc, Steele:1981tn}, we obtain
the stochastic differential equations
\begin{equation}
\label{conres}
\begin{gathered}
dR=\left(\alpha R\left(1-\frac{R}{\beta}\right)-\frac{\delta R^{2}C}{\kappa+R^{2}}\right)dt+\sigma_{1}dW_{1}\\[5pt]
dC=\left(\frac{\gamma R^{2}C}{\kappa+R^{2}}-\mu C^{2}\right)dt+\sigma_{2}dW_{2} .
\end{gathered}
\end{equation}
Here $W_{1}$ and $W_{2}$ are independent Wiener processes. The resource has logistic growth in the absence of consumers, with maximum
growth rate $\alpha$ and carrying capacity $\beta$. Consumption
of resources is represented by a sigmoidal Type III functional response.
$\delta$ is the
maximum consumption rate, and $\kappa$ controls
how quickly the consumption rate saturates. $\gamma$ determines the conversion from resources to consumers. The consumers
have a quadratic mortality term with coefficient $\mu$, which represents
the negative impacts of intraspecific competition. $\sigma_{1}$ and
$\sigma_{2}$ are the noise intensities for the resource and consumer
populations, respectively.

The additive form of the stochastic terms in this model represent random inputs
and losses of resources and consumers. In situations where inherent
growth parameters (e.g., $\alpha$ or $\gamma$) are stochastic, other
forms of stochasticity would be appropriate. We will deal with additive noise here; the more general case is considered in appendix~\ref{subsec:ONS}.

\begin{figure}[ht]
\centering
\begin{subfigure}[b]{0.4\textwidth}
\caption{}
\includegraphics[width=\textwidth]{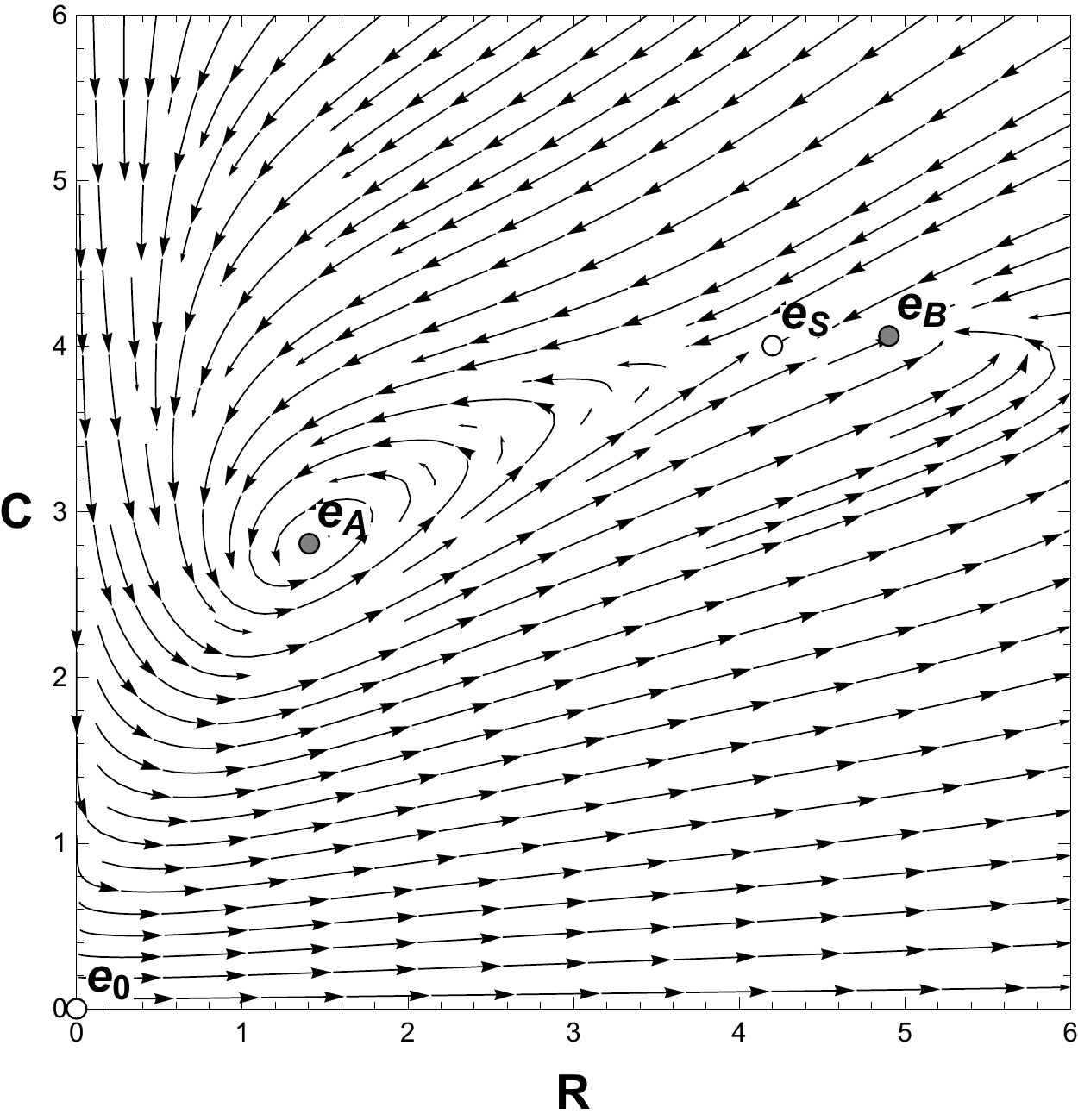}
\end{subfigure}
\begin{subfigure}[b]{0.4\textwidth}
\caption{}
\includegraphics[width=\textwidth]{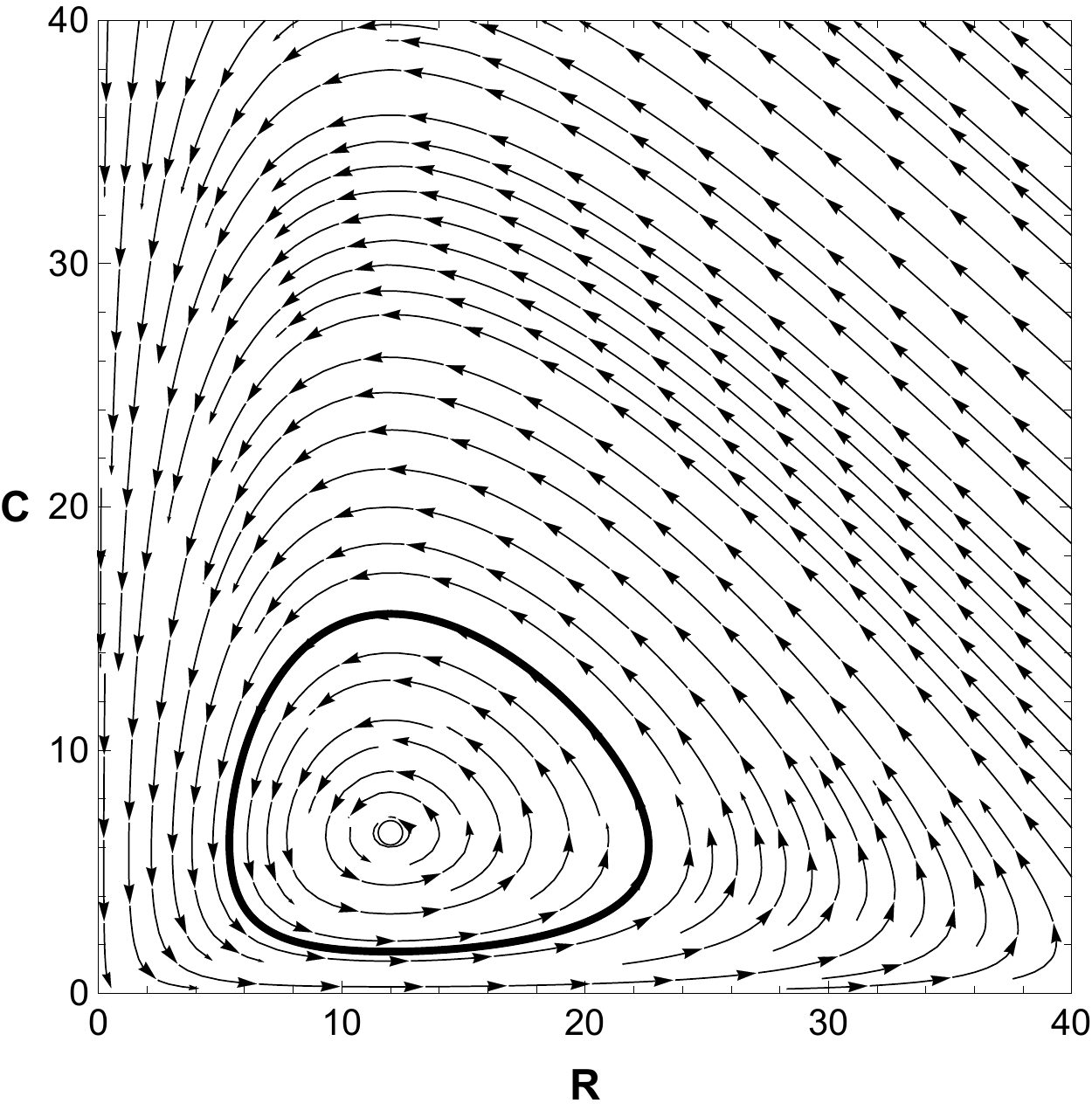}
\end{subfigure}
\vspace*{3mm}
\begin{subfigure}[b]{0.4\textwidth}
\caption{}
\includegraphics[width=\textwidth]{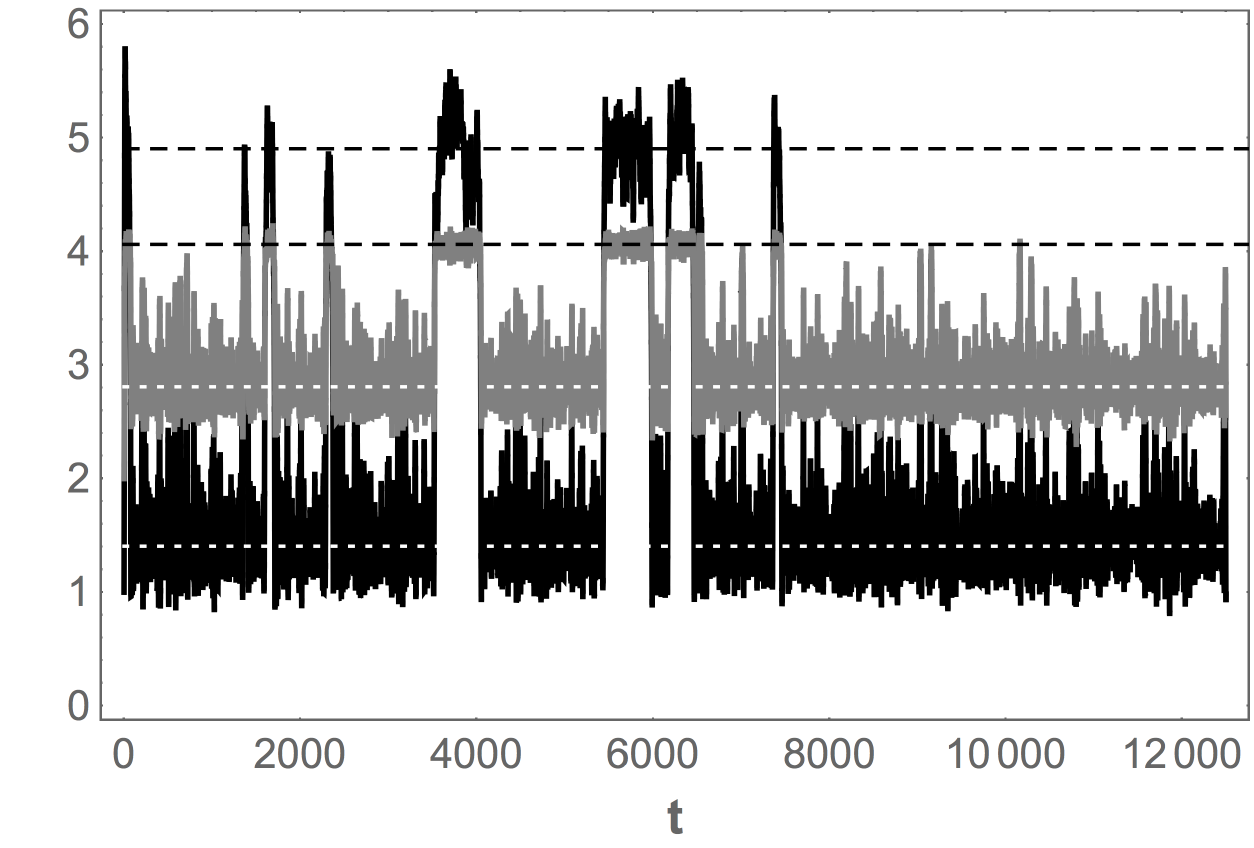}
\end{subfigure}
\begin{subfigure}[b]{0.4\textwidth}
\caption{}
\includegraphics[width=\textwidth]{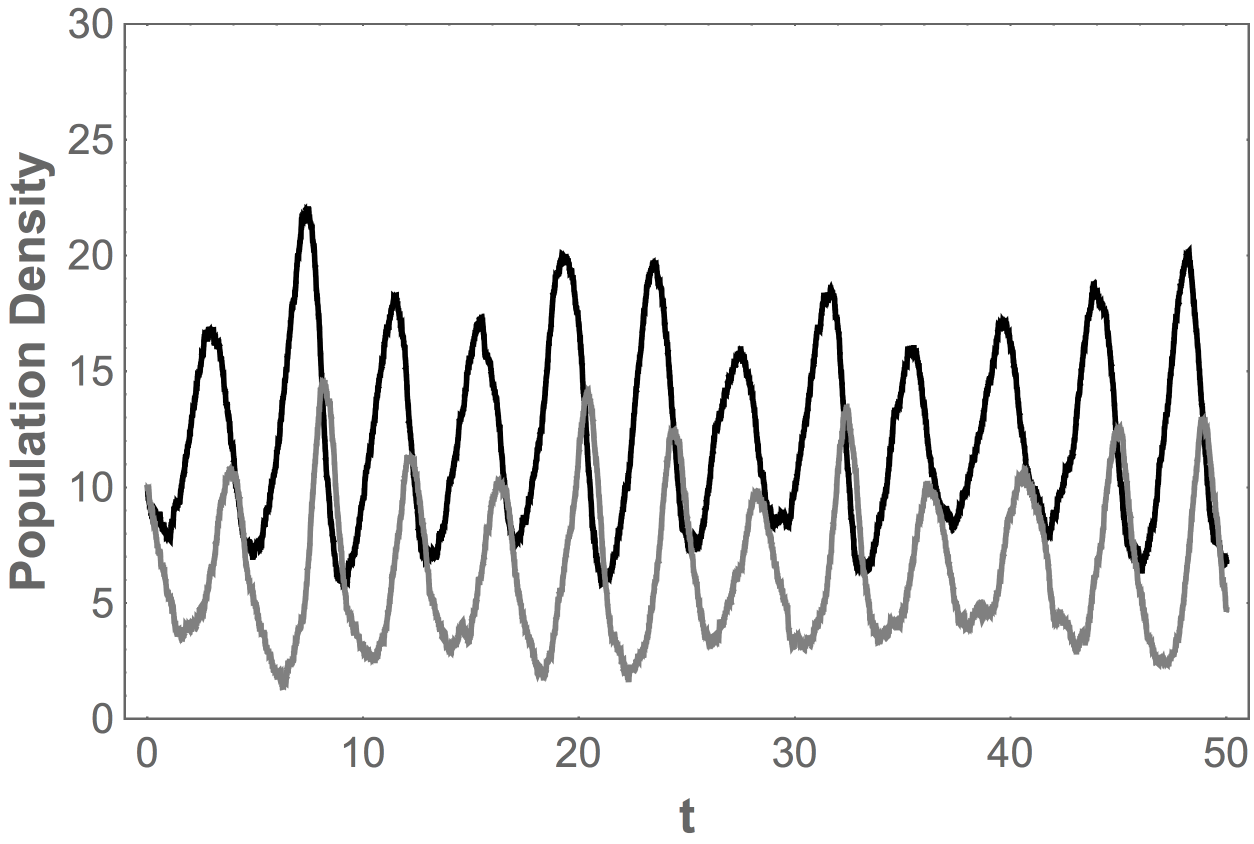}
\end{subfigure}
\caption{\small {\label{Fig2} \textbf{(a)} Stream plot for the deterministic skeleton of the consumer-resource model in example 2. Unstable equilibria are white disks and stable equilibria are gray disks. The unstable equilibrium $\mathbf{e}_P$ is not shown, but would appear on the x-axis to the right of where the graph is truncated. Variables are scaled, so the units are dimensionless. Lines and arrows show the direction of trajectories for equations \eqref{conres} in the absence of noise. \textbf{(b)} Similar stream plot for the deterministic skeleton of  the consumer-resource model in example 3. The white disk is an unstable equilibrium and the gray line is a stable limit cycle. \textbf{(c)} A realization of equations~\eqref{conres} for example 2. Integration was performed with the Euler-Maruyama method and $\Delta t=0.025$. Resource population density is black and consumer population density is gray. The dotted white lines correspond to the equilibrium $\mathbf{e}_{A}$, and the dashed black lines to the equilibrium $\mathbf{e}_{B}$. \textbf{(d)} A realization of equations~\eqref{predprey} for example 3, with $\sigma=0.8$. Integration was performed with the Euler-Maruyama method and $\Delta t=5\times10^{-4}$. Resource population density is black and consumer population density is gray.}}
\end{figure}

We will analyze \eqref{conres} with parameters set at $\alpha\!=\!1.54$, $\beta\!=\!10.14$,
$\gamma\!=\!0.476$, $\delta\!=\!\kappa\!=\!1$, and $\mu\!=\!0.112509$. A phase plot of the deterministic skeleton is shown in figure~\ref{Fig2}a. The deterministic skeleton
of this system has five equilibria: $\mathbf{e}_{0}\!=\!(0,0)$,
$\mathbf{e}_{A}\!=\!(1.405,2.808)$,
$\mathbf{e}_{B}\!=\!(4.904,4.062)$,
$\mathbf{e}_{S}\!=\!(4.201,4.004)$,
$\mathbf{e}_{P}\!=\!(\beta,0)$.

A linear stability analysis shows that $\mathbf{e}_{0}$ is an unstable equilibrium and $\mathbf{e}_{P}$ is a saddle point. $\mathbf{e}_{A}$ and $\mathbf{e}_{B}$ are stable equilibria, and $\mathbf{e}_{S}$ is a saddle point that lies between them. Equilibria and their stability are summarized in figure~\ref{Fig2}a.

The eigenvalues of the Jacobian are $-0.047\pm0.458i$ at  $\mathbf{e}_{A}$ and $-0.377$ and $-0.093$ at $\mathbf{e}_{B}$. For $\mathbf{e}_{A}$ the real part of the eigenvalue with largest real part is $-0.047$, and for $\mathbf{e}_{B}$ it is $-0.093$; therefore, the stability analysis
concludes that $\mathbf{e}_{B}$ is more stable, because this
value is more negative than it is for $\mathbf{e}_{A}$.

A realization of the stochastic system ($\sigma_{1}=\sigma_{2}=0.05$, Figure~\ref{Fig2}c) shows switching between the two stable states. It
is typical of most realizations we generated, in that it spends more
time near $\mathbf{e}_{A}$ (dotted white lines) than $\mathbf{e}_{B}$ (dashed black lines). This realization, which had initial condition $\left(x_{0},\,y_{0}\right)=\left(1,\,2\right)$, spent 87\% of its time in the basin of attraction corresponding to $\mathbf{e}_{A}$. Intuitively,
it seems that $\mathbf{e}_{A}$ should be classified as more stable
than $\mathbf{e}_{B}$, but as in Example 1, this is not what was obtained via
the standard linear stability analysis.

Recall that realizations are of limited utility for stability analysis, because each value of $\sigma$ will produce different dynamics and different steady\-state probability distributions (see appendix D and supplementary figure 1). The potential is defined independently of $\sigma$, and hence would be ideal for providing more general insights than $\sigma$-specific realizations.  Of course, we do not have a potential function $U$ for this or any other non\-gradient system and
hence cannot compare $U(\mathbf{e}_{A})$ and $U(\mathbf{e}_{B})$.
Instead, we turn to the Freidlin\-Wentzell quasi\-potential, which generalizes
the notion of a potential.
\medskip
\section*{Generalizing The Potential}

For higher\-dimensional
models, we need to introduce a little bit of new notation.  We can write an $n$\-dimensional system of stochastic differential equations with additive noise as
\begin{equation}
\label{gradient3}
d\mathbf{X}=f(\mathbf{X})\,dt+\sigma\,d\mathbf{W}.
\end{equation}
$\mathbf{X}\!=\!\left(X_{1},\ldots,X_{n}\right)$ is a column vector of
state variables and $\mathbf{W}\!=\!\left(W_{1},\ldots,W_{n}\right)$ is
a column vector of $n$ independent Wiener processes. We use the lowercase notation $\mathbf{x}\!=\!\left(x_{1},\ldots,x_{n}\right)$ to indicate a point in phase space (as opposed to a stochastic process). $f$
is the deterministic skeleton of the system. It is a vector field: for every point $\mathbf{x}$, $f\!(\mathbf{x})$ specifies the direction that a deterministic trajectory will move. $\sigma$ is the noise intensity. More general ways of incorporating noise are considered in appendix~\ref{subsec:ONS}.

Following the same
general approach as in example 1, the Fokker\-Planck equation
for a two dimensional version of~\eqref{gradient3}, with $\mathbf{X}\!=\!\left(X_{1},X_{2}\right)$, $\mathbf{x}\!=\!\left(x_{1},x_{2}\right)$ and $f\!=\!\left(f_{1},f_{2}\right)$, is: 
\begin{equation}
\label{FP}
\frac{\partial p}{\partial t}=-\frac{\partial}{\partial x_{1}}\left(f_{1}p\right)-\frac{\partial}{\partial x_{2}}\left(f_{2}p\right)+\frac{\sigma^{2}}{2}\left(\frac{\partial^{2}p}{\partial x_{1}^{2}}+\frac{\partial^{2}p}{\partial x_{2}^{2}}\right).
\end{equation}
In the gradient case in Example 1, the steady\-state solution of the Fokker\-Planck equation was of the form
\eqref{steadystate} (replacing $x$ with $\mathbf{x}$ and obtaining $Z$ via integration over the positive quadrant). Here, there is no function $U$ to play that role, but using
the same general approach, assume that there is a function $V(\mathbf{x})$
such that: 
\begin{equation}
\label{effecpot}
p_{s}(\mathbf{x})\asymp\exp\left(-\frac{2V(\mathbf{x})}{\sigma^{2}}\right).
\end{equation}
The symbol $\asymp$ denotes logarithmic equivalence, details about which are in appendix~\ref{subsec:FWAppendix}. When noise intensity is small, we can obtain an approximation for $V$ (using asymptotic expansion; see appendix~\ref{subsec:SmallNAppendix}). This approximation, denoted by $V_{0}(\mathbf{x})$, satisfies
\begin{equation}
\label{HJE}
\nabla V_{0}\cdot\nabla V_{0}+f\cdot\nabla V_{0}=0,
\end{equation}
where the gradient operator $\nabla$ takes a scalar function $\psi$ as an input, and
returns a vector,
$\nabla\psi\!=\!\left(\frac{\partial\psi}{\partial x_{1}},\frac{\partial\psi}{\partial x_{2}},\ldots,\frac{\partial\psi}{\partial x_{n}}\right)$, that is the multi\-dimensional analogue of the derivative.
Intuitively, if one thinks of $\psi(\mathbf{x})$ as specifying the height of a landscape at a particular point $\mathbf{x}$, then 
$-\nabla\psi(\mathbf{x})$ points in direction of the steepest descent (as water would flow).

Equation \eqref{HJE} is the static Hamilton\-Jacobi equation. Interestingly, $V_{0}$ has key properties that make it a useful analog of a potential in a gradient system. First, $V_{0}$ is independent
of the noise intensity $\sigma$, just as the potential function $U$
was in the gradient case. Second, if $\mathbf{x}(t)$ is trajectory
of the deterministic skeleton of \eqref{gradient3}, then
\begin{equation}
\frac{d}{dt}\left(V_{0}\left(\mathbf{x}(t)\right)\right)=\nabla V_{0}\cdot f\left(\mathbf{x}(t)\right)=-\nabla V_{0}\cdot\nabla V_{0}\leq0,
\end{equation}
and $\frac{d}{dt}\left(V_{0}\left(\mathbf{x}(t)\right)\right)=0$
only where $\nabla V_{0}=0$. Thus $V_{0}$ is a Lyapunov function
for the deterministic system, which is an important feature for the ball\-in\-cup metaphor. If $V_{0}(\mathbf{x})$ specifies an two\-dimensional surface, then, in the absence of perturbations,
trajectories will always move ``downhill''. Again, this parallels
the role that $U$ played in the gradient systems. Third, we can interpret the relationship between $f$ and the surface $V_{0}$. $f$ is the deterministic skeleton that causes trajectories to move across the landscape, and $-\nabla V_{0}$ is the component of $f$ that causes trajectories to move downhill. The remaining component of $f$, which we denote by $Q$ and call the ``circulatory" component, is defined as:
\begin{equation}
Q\left(\mathbf{x}\right)=f\left(\mathbf{x}\right)+\nabla V_{0}\left(\mathbf{x}\right).
\end{equation}
$V_{0}$ satisfies the Hamilton\-Jacobi equation, so
$Q\cdot\nabla V_{0}\!=\!f\cdot\nabla V_{0}+\nabla V_{0}\cdot\nabla V_{0}=0$,  hence $\nabla V_{0}$ and $Q$ are perpendicular at every point. This motivates the label ``circulatory" -- in the absence of other forces, $Q$ would cause trajectories to circulate around level sets of $V_{0}$.

The function $V_{0}$ generalizes the potential function to non\-gradient systems and extends to $n$\-dimensional systems. Interestingly, $V_{0}$ is
a scalar multiple of a function called the Freidlin\-Wentzell quasi\-potential.
The quasi\-potential has extremely important properties, which we explore
in the next section before applying all of these ideas to example~2.
\medskip
\section*{The Freidlin\-Wentzell Quasi\-potential}

Freidlin and Wentzell \citeyearpar{Freidlin:2012wd} analyzed stochastic differential equations using
a large deviation principle, which is an asymptotic law determining the probabilities
of different trajectories. These concepts can be best interpreted by imagining the state of the system (the position of the ball, or the current combination of population densities) being randomly perturbed within a ``force field'' imposed by the deterministic skeleton. Suppose the system starts at the stable state $\mathbf{e}_{A}$
and travels to another state $\mathbf{x}$. To complete this journey, the
populations will need to do some ``work'' against the force field (i.e., they need to go ``uphill");
this work is provided by random perturbations. Trajectories
that require the least amount of work (require the least extreme
stochastic perturbations) are the most likely. Suppose that $\mathbf{\theta}\!\left(t\right)$ specifies a path, parameterized by $t$,
that goes from the stable equilibrium $\mathbf{\theta}(0)=\mathbf{e}_{A}$ to another state $\mathbf{\theta}(T)=\mathbf{x}$. $T$ is total time it takes the populations to move along this path from $\mathbf{e}_{A}$ to $\mathbf{x}$. The amount of work required for the populations to follow a given path can be quantified by a functional $S_{T}$ called the action 
(see appendix~\ref{subsec:FWAppendix} for details). 

In order to determine the amount of work it takes to get to some state $\mathbf{x}$, one must minimize the action over all possible paths from $\mathbf{e}_{A}$ to $\mathbf{x}$, and all path durations $T\!>\!0$. The minimum action is called the quasi\-potential, denoted $\Phi_{\mathbf{e}_{A}}\!(\mathbf{x})$. The quasi\-potential depends on the starting point $\mathbf{e}_{A}$; when there are multiple stable states, the corresponding quasi-potentials can be stitched together to obtain a global quasi\-potential, $\Phi(\mathbf{x})$ \citep{roy1994}; see further details in appendix~\ref{subsec:GQP}. $\Phi$ is related to $V_{0}$ by $\Phi=2\,V_{0}$ (appendix~\ref{subsec:HJEQP}). In this paper, we use $V_{0}$ instead of $\Phi$, because $V_{0}$ agrees with the true potential in gradient systems. The multiple of $2$ in the relationship $\Phi=2\,V_{0}$ is an inconvenient result of the Freidlin\-Wentzell definition. Conceptually, these two functions measure the same properties, and computing one immediately yields the other.

The quasi\-potential can be calculated by solving the static
Hamilton\-Jacobi equation~\eqref{HJE}. This is a numerically difficult task,
however; standard finite difference and finite element methods typically
break down when applied to this kind of non\-linear partial differential equation.
Ordered upwind methods \citep{Sethian:2001wx} are an innovative approach that
circumvent the problems encountered by traditional methods. The basic
idea is to create an expanding front of points where the solution
is known, and march outward by considering and accepting solution values at adjacent
points in ascending order. For use in systems of the form~\eqref{gradient3}, the
standard ordered upwind method was enhanced by \citet{Cameron:2012ex}.
Cameron's algorithm allows for efficient computation of the quasi\-potential. It forms the basis for \textit{QPot}, a freely\-available R package we have developed \citep{Moore2015} that includes a full set of tools for analyzing two-dimensional autonomous stochastic differential equations. \textit{QPot} can be downloaded at CRAN \footnote{The Comprehensive R Archive Network, https://cran.r-project.org} or GitHub \footnote{https://github.com/bmarkslash7/QPot}. To calculate the quasi-potential, users simply input the deterministic skeleton of the system, the domain, and the mesh size (although many other options are available). Computation time for the ordered upwind method depends on the model and mesh size; example~2 took less than ten minutes on a fairly average personal computer.

The Freidlin-Wentzell construction of the quasi-potential provides a mathematically rigorous justification for the Wentzel-Kramers-Brillouin (WKB) ansatz, which can be used to approximate mean first passage times in the small noise limit \citep{Bressloff:2014dx}. The WKB method has been applied to calculate expected extinction times for several specific models in population dynamics and epidemiology \citep{Meerson:2009bv, Roozen:1989cs, vanHerwaarden:1995el, Ovaskainen:2010cla}.
\medskip
\section*{Example 2 Continued}

We generated solutions to the static Hamilton\-Jacobi equation for the system~\eqref{conres} using base points $\mathbf{e}_{A}$ and
$\mathbf{e}_{B}$, and then matched them into a global quasi\-potential
by enforcing continuity at $\mathbf{e}_{S}$ and setting the minimum to 0. We divided this function by two to obtain $V_{0}$. The ordered upwind method was implemented using Cameron's algorithm \citep{Cameron:2012ex}. \textit{Mathematica} was used for data processing and graphics generation, and the code is available as a supplementary file.

For the consumer\-resource
system \eqref{conres}, the resulting surface for $V_{0}$ and a corresponding contour plot are shown in figure~\ref{Fig4}a-b. We find that $V_{0}(\mathbf{e}_{A})\!=\!0$,
$V_{0}(\mathbf{e}_{S})\!=\!0.007$, $V_{0}(\mathbf{e}_{B})\!=\!0.006$.
The relative values of $V_{0}$ can be used to make calculations regarding
first passage times and calculate transition rates between $\mathbf{e}_{A}$
and $\mathbf{e}_{B}$. The most fundamental observation, however,
is that $V_{0}(\mathbf{e}_{A})\!<\!V_{0}(\mathbf{e}_{B})$,
which indicates that $\mathbf{e}_{A}$ is more stable than $\mathbf{e}_{B}$.
This contrasts with the linear stability analysis, but agrees with the
qualitative picture obtained from realizations of the system. As in
example 1, analyzing the system through the lens of a potential
(or quasi\-potential) function yields a completely different conclusion
than the deterministic analysis, and one that aligns much more clearly with the simulated dynamics we observe. Furthermore, $V_{0}(\mathbf{e}_{S})$
and $V_{0}(\mathbf{e}_{B})$ are closer to each other than
they are to $V_{0}(\mathbf{e}_{A})$. This indicates that
$\mathbf{e}_{S}$ and $\mathbf{e}_{B}$ have similar stabilities, and it encourages us to move beyond the dichotomous classification of equilibria as either stable or unstable, which is often applied in linear stability analysis. The stable vs.~unstable dichotomy
classifies $\mathbf{e}_{A}$ and $\mathbf{e}_{B}$ as alike,
and $\mathbf{e}_{S}$ as different. The quasi\-potential shows
that it is $\mathbf{e}_{B}$ and $\mathbf{e}_{S}$ that are
alike, and $\mathbf{e}_{A}$ that is different. By quantifying
stability on a useful continuum, the quasi\-potential offers a more nuanced
perspective.

$V_{0}$ also provides a useful way to decompose the deterministic skeleton of equations~\eqref{conres} into physically interpretable parts, $f=-\nabla V_{0}+Q$. This decomposition is shown in figure~\ref{Fig5}a-b. $-\nabla V_{0}$ represents the part of the system that moves the system towards stable states, while $Q$ represents the part that causes consumer\-resource cycling.
\medskip
\section*{Example 3: Predator and Prey With A Limit Cycle}

The quasi\-potential allows for stability analysis of attractors that
are more complicated than equilibrium points. As discussed in \citet{Cameron:2012ex} and \citet{Freidlin:2012wd} and explained in appendix~\ref{subsec:FWAppendix}, the quasi\-potential can be defined for compact sets, such as limit cycles. As an example
of a non\-gradient system with a limit cycle, consider a stochastic version of the Rosenzweig\-MacArthur predator\-prey model \citep[e.g.][]{logan2009mathematical}:
\begin{equation}
\label{predprey}
\begin{gathered}
dR=\left(\alpha R\left(1-\frac{R}{\beta}\right)-\frac{\delta RC}{\kappa+R}\right)dt+\sigma_{1}dW_{1}\\[5pt]
dC=\left(\frac{\gamma RC}{\kappa+R}-\mu C\right)dt+\sigma_{2}dW_{2} .
\end{gathered}
\end{equation}
Here $R$ is the resource density, $C$ is the consumer density,
and $W_{1}$ and $W_{2}$ are independent Wiener processes. Consumption of resources is represented by a Type II functional response; otherwise the resource dynamics are the same as  in example 2. In the absence of resources, the consumer density decreases at an exponential
rate determined by $\mu$. $\sigma_{1}$ and $\sigma_{2}$ are the
noise intensity for the resource and consumer densities, respectively.
We present the analysis of this model with $\alpha\!=\!1.5$, $\beta\!=\!45$,
$\gamma\!=\!5$, $\delta\!=\!10$, $\kappa\!=\!18$, and $\mu\!=\!4$.

Figure~\ref{Fig2}b,d shows a stream plot of the system's deterministic skeleton, and a realization with noise intensities $\sigma_{1}\!=\!\sigma_{2}\!=\!0.8$ over time interval $[0,50]$. This choice of noise intensity and time scale was made to illustrate clear population cycles with amplitude shifts.

\begin{figure}[ht]
\centering
\begin{subfigure}[b]{0.4\textwidth}
\caption{}
\includegraphics[width=\textwidth]{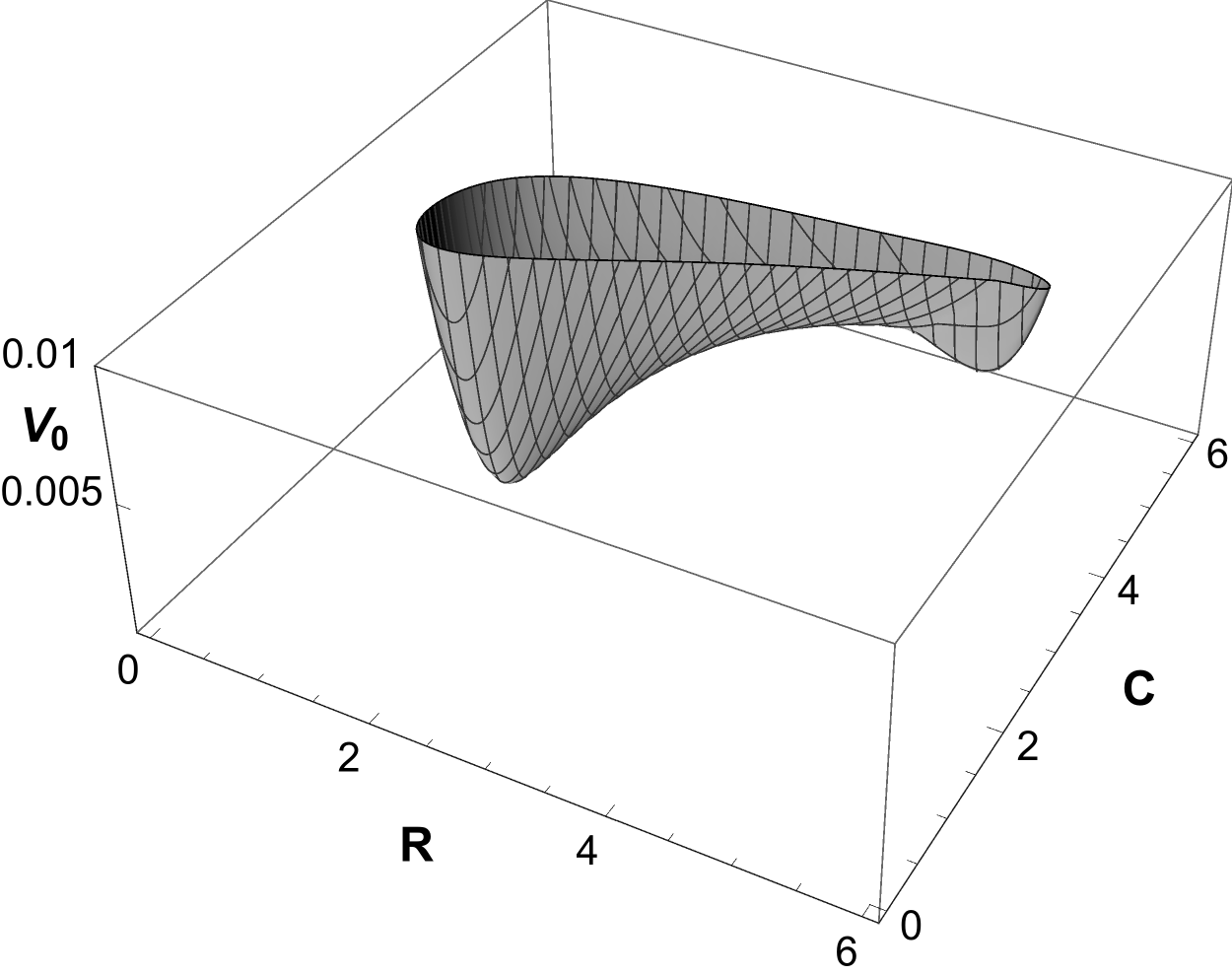}
\end{subfigure}
\begin{subfigure}[b]{0.4\textwidth}
\caption{}
\includegraphics[width=\textwidth]{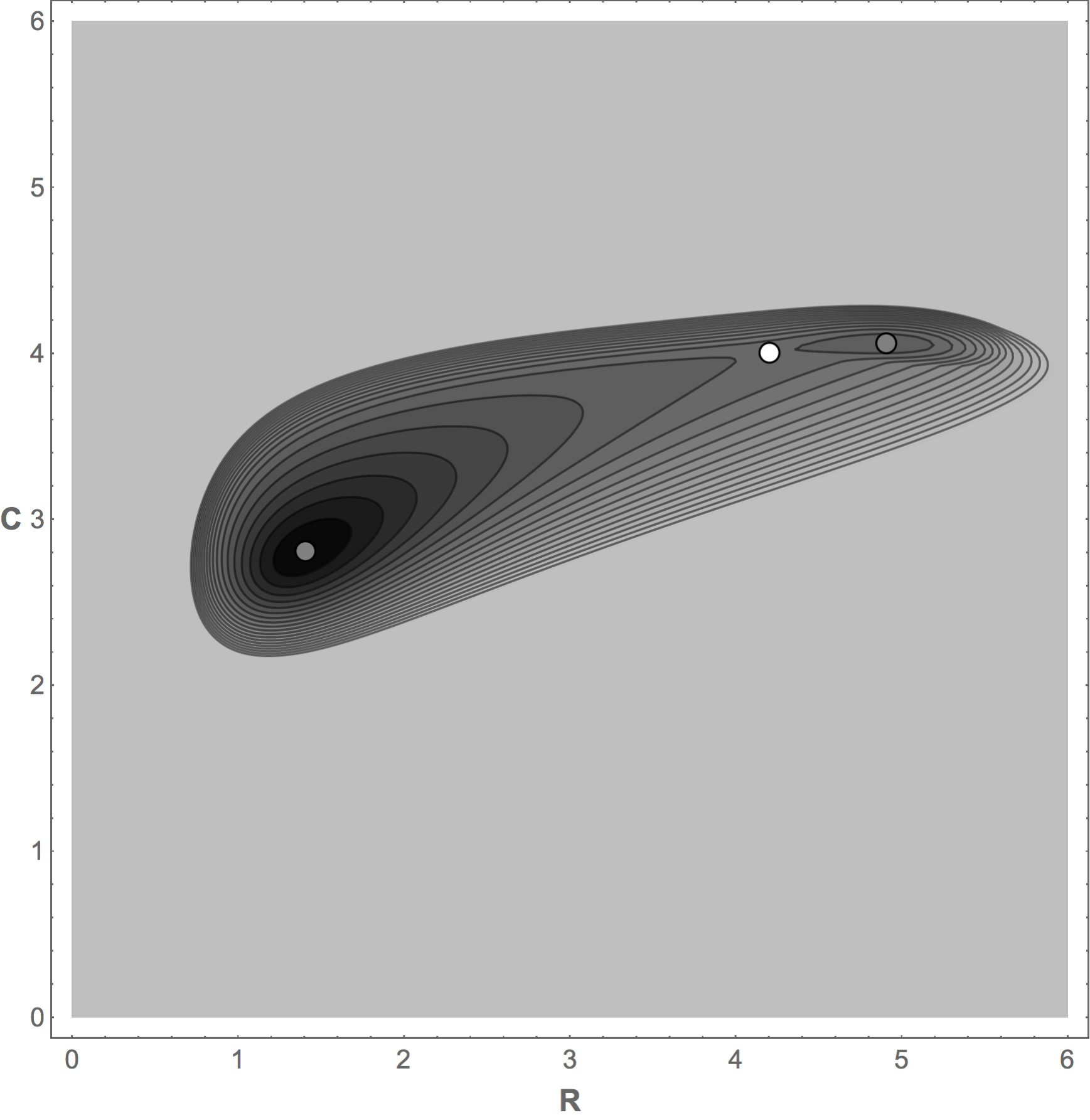}
\end{subfigure}
\vspace*{3mm}
\begin{subfigure}[b]{0.4\textwidth}
\caption{}
\includegraphics[width=\textwidth]{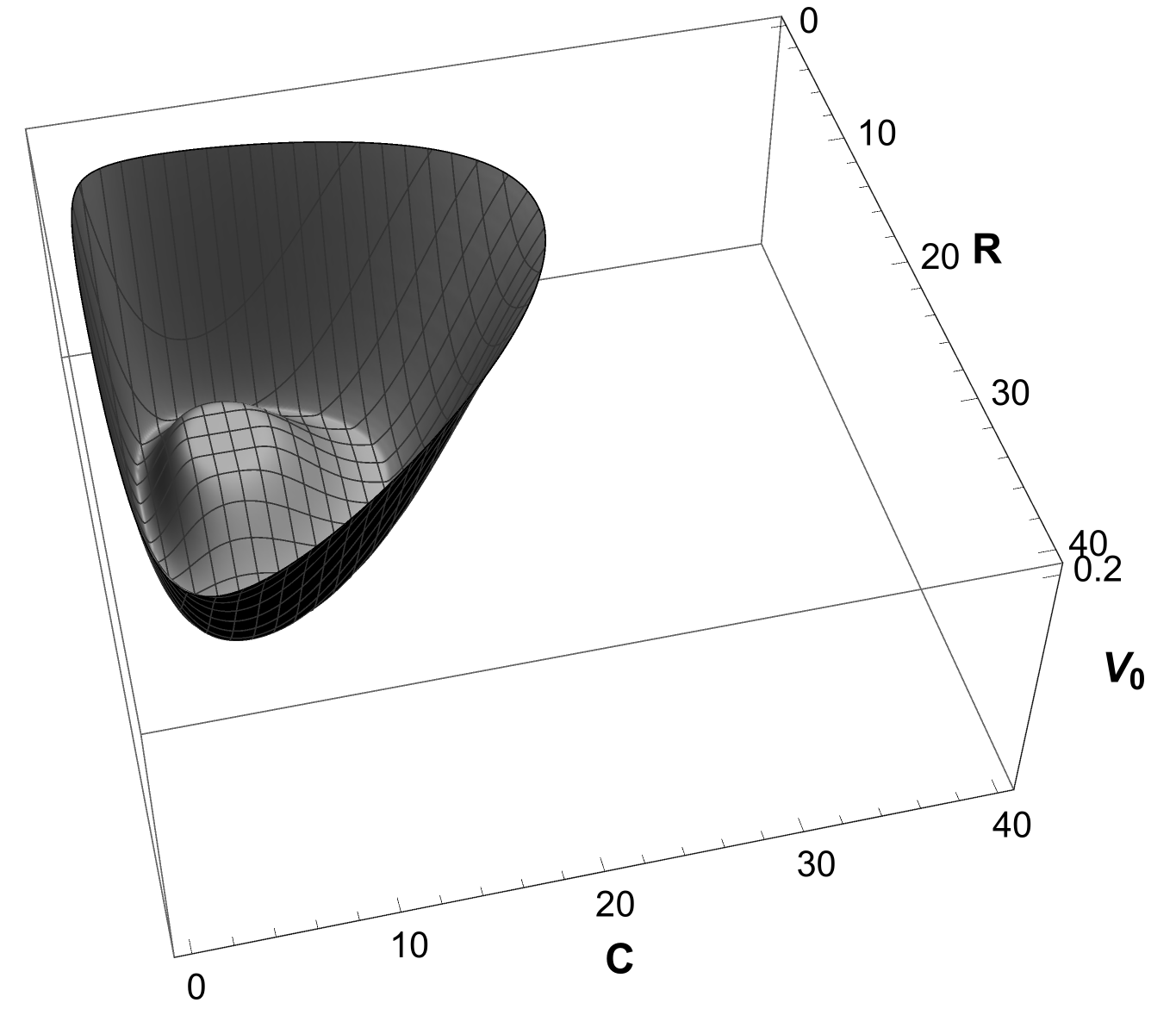}
\end{subfigure}
\begin{subfigure}[b]{0.4\textwidth}
\caption{}
\includegraphics[width=\textwidth]{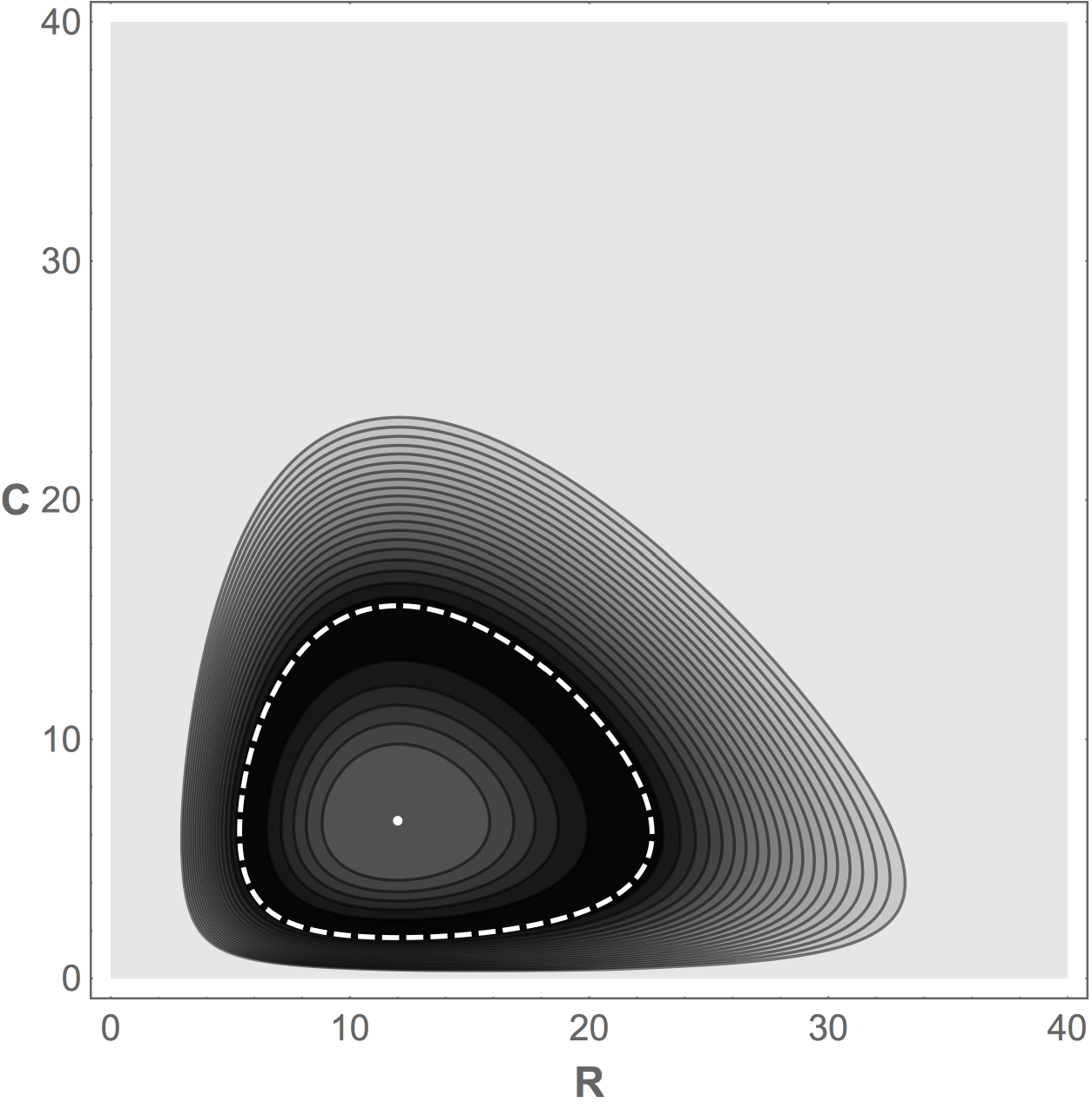}
\end{subfigure}
\caption{\small {\label{Fig4} \textbf{(a)} The quasi-potential function for the consumer-resource model, equations~\eqref{conres}. Variables are scaled, so the units are dimensionless. Note that the quasi-potential surface is much deeper around $\mathbf{e}_{A}$ than $\mathbf{e}_{B}$. The quasi-potential is truncated at 0.02 for display purposes; it continues to increase in the regions outside the plot. \textbf{(b)} Contour plot for the same model. The white disk is the saddle point $\mathbf{e}_{S}$. The gray disks are the stable equilibria $\mathbf{e}_{A}$ and $\mathbf{e}_{B}$. \textbf{(c)} The quasi-potential function for equations~\eqref{predprey}. \textbf{(d)} Contour plot for the same model. The white disk is an unstable equilibrium, and the white dashed line is a stable limit cycle.}}
\end{figure}

\begin{figure}[ht]
\centering
\begin{subfigure}[b]{0.4\textwidth}
\caption{}
\includegraphics[width=\textwidth]{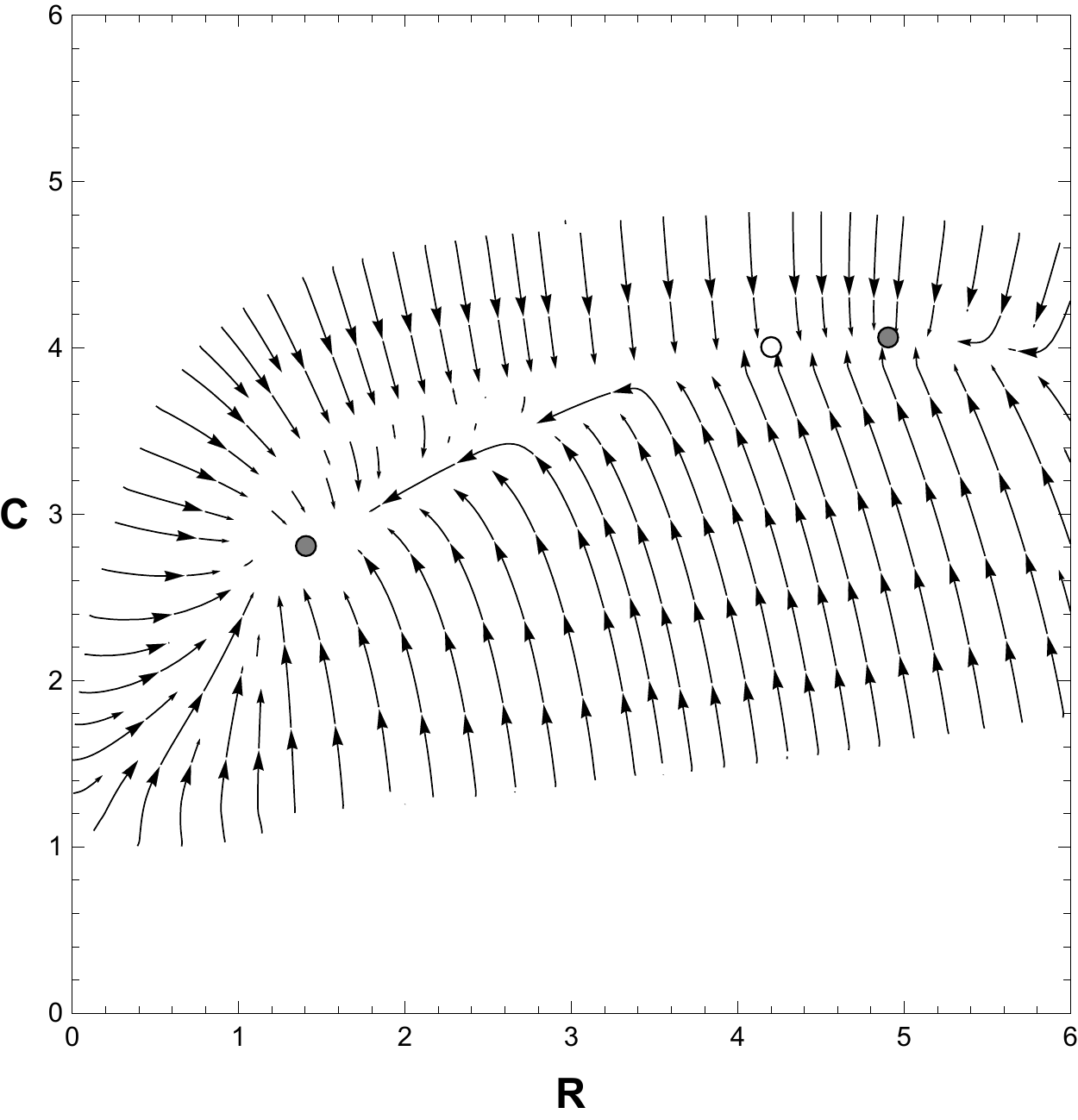}
\end{subfigure}
\begin{subfigure}[b]{0.4\textwidth}
\caption{}
\includegraphics[width=\textwidth]{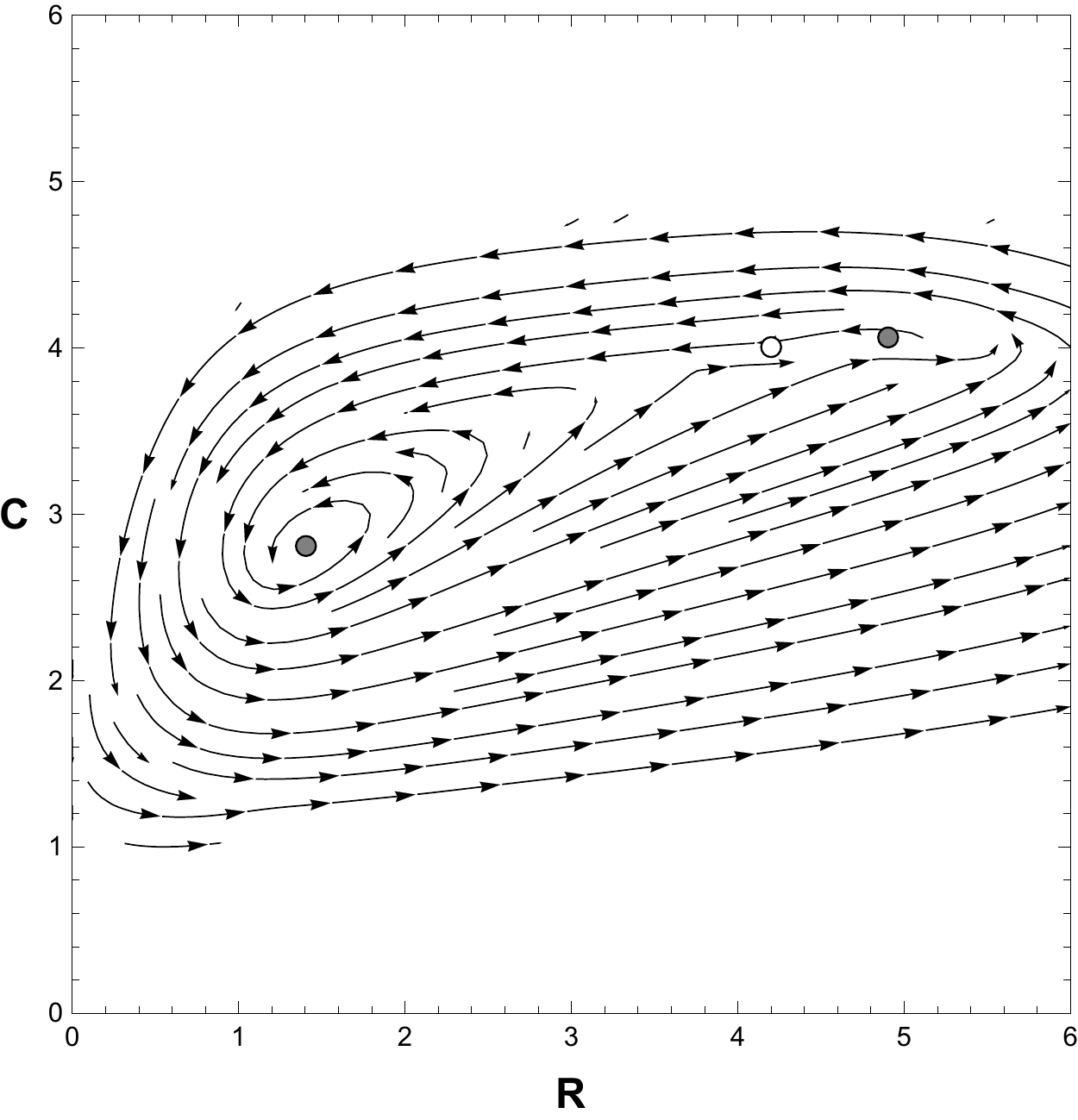}
\end{subfigure}
\vspace*{3mm}
\begin{subfigure}[b]{0.4\textwidth}
\caption{}
\includegraphics[width=\textwidth]{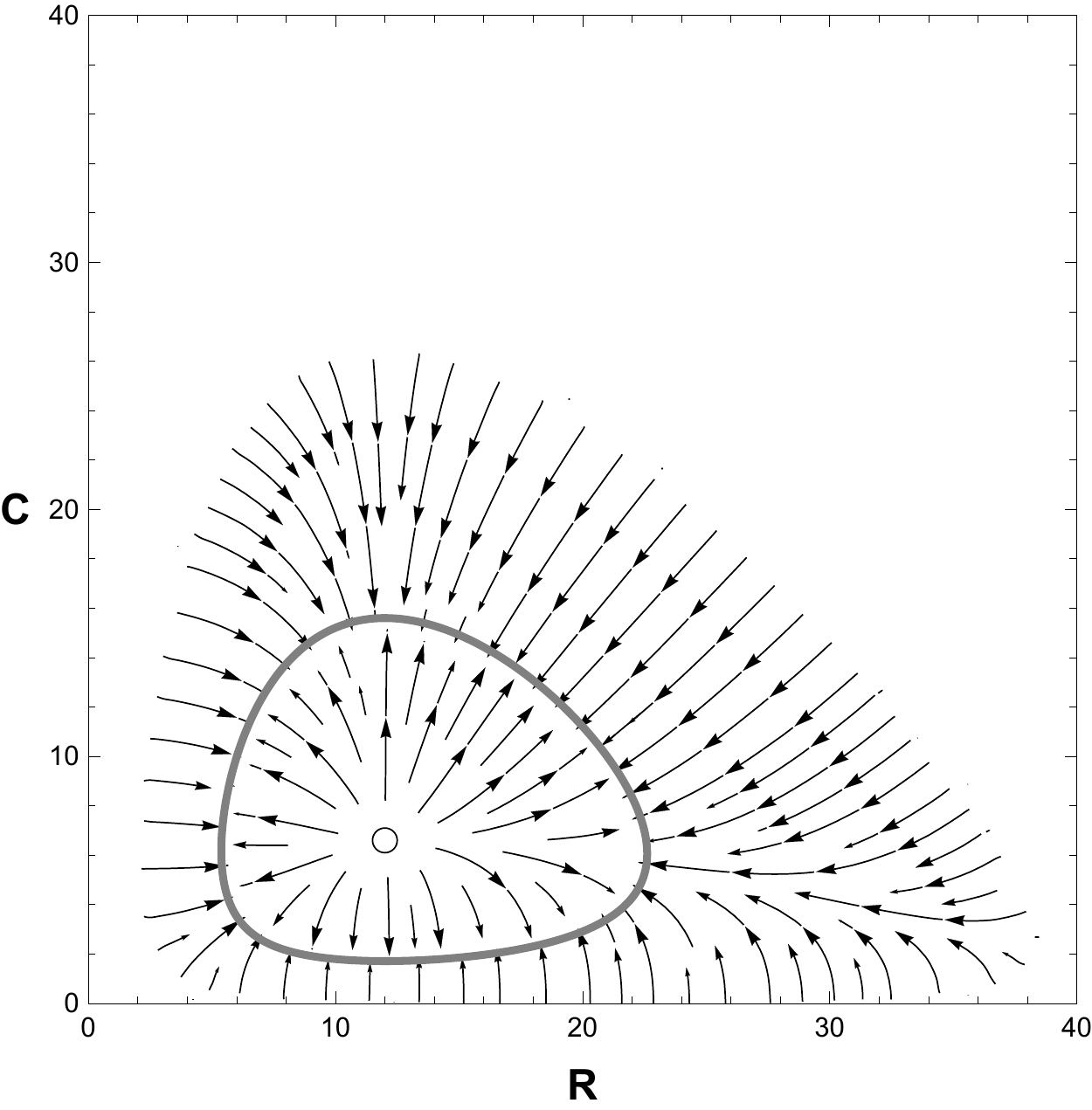}
\end{subfigure}
\begin{subfigure}[b]{0.4\textwidth}
\caption{}
\includegraphics[width=\textwidth]{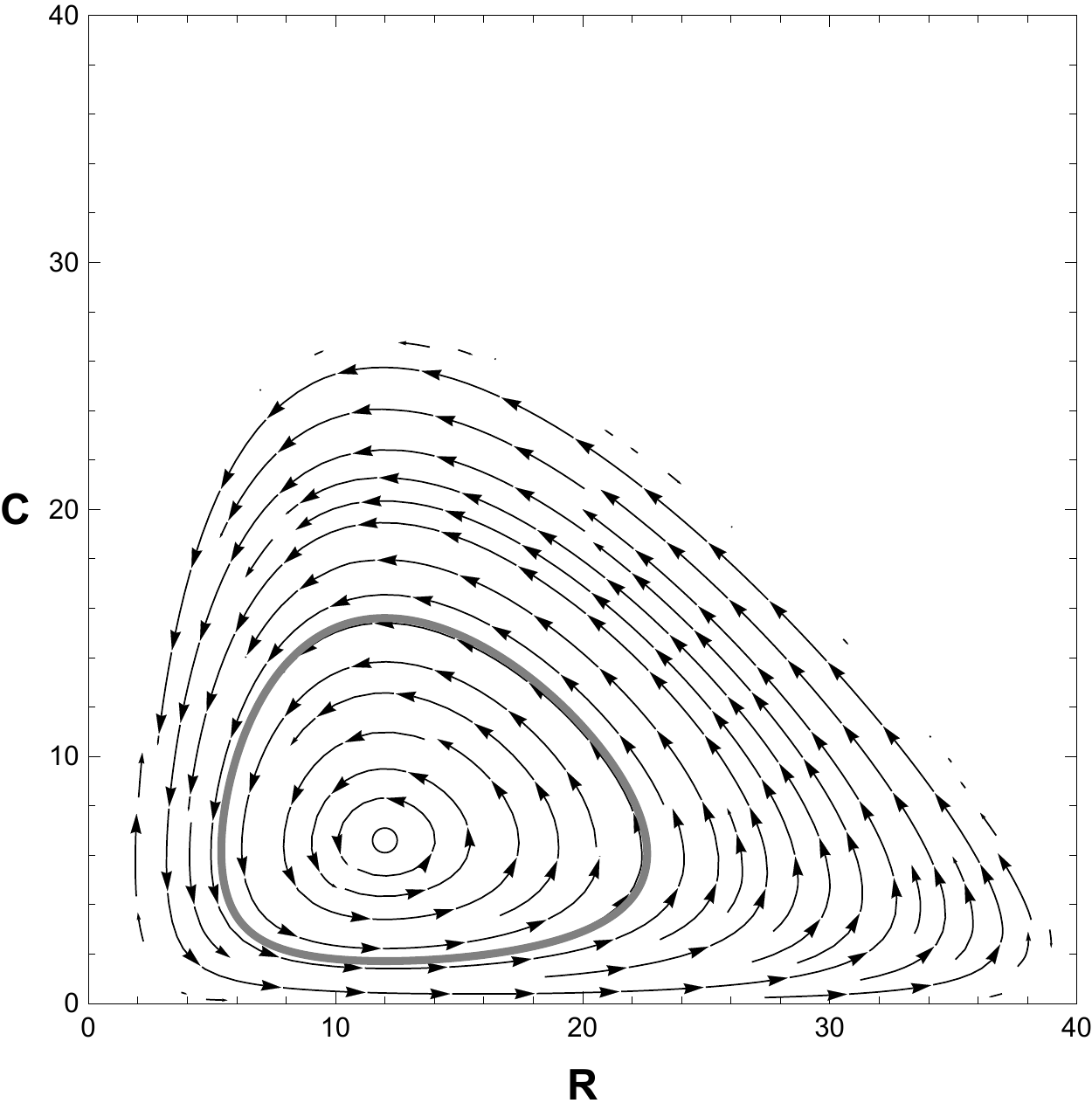}
\end{subfigure}
\caption{\small {\label{Fig5} \textbf{(a)} and \textbf{(b)} are the orthogonal decomposition of the deterministic skeleton of the system~\eqref{conres}. \textbf{(a)} The ``downhill" component, $-\nabla V_{0}$. \textbf{(b)} The ``circulatory" component, $Q$. Gray disks are stable equilibria. The white disk is an unstable equilibrium. \textbf{(c)} and \textbf{(d)} are the orthogonal decomposition of the deterministic skeleton of the system~\eqref{conres}. The thick gray line is a stable limit cycle.}}
\end{figure}

Surface and contour plots of $V_{0}$ for system~\eqref{predprey} are shown in figure~\ref{Fig4}c-d. Recall that $V_{0}$ provides
a decomposition of the deterministic system into a ``downhill" force and a ``circulatory" force, as shown in figure~\ref{Fig5}c-d. In this case, $-\nabla V_{0}$ causes trajectories to be attracted to the limit
cycle's trough. The circulatory component causes trajectories to cycle
in this trough. This decomposition harkens back to \citet{Holling:1973wh}, who made the following observation about dynamical
systems: ``There are two components that are important:
one that concerns the cyclic behavior and its frequency and amplitude,
and one that concerns the configuration of forces caused by the positive
and negative feedback relations." The latter is described by the
gradient of $V_{0}$, the former by the circulatory component.
Therefore, we see that the Freidlin\-Wentzell approach provides a systematic
way to distinguish between the two concepts identified by Holling.

In this example, we cannot contrast the quasi\-potential results
with the traditional linear stability analysis, because the latter only applies to equilibrium points.
\medskip
\section*{Limitations and Generalizations}

In this paper, we have focused on applying the quasi-potential framework to stochastic differential equations models that share several characteristics: 1) time is continuous, 2) state variables are continuous, 3) noise is additive and the noise intensity is the same for both state variables, 4) noise is a direct perturbation to the state variables (as opposed to a perturbation to parameter values), 5) noise is white (as opposed to colored), and 6) noise occurs continually with low intensity (as opposed to occurring as discrete, abrupt events). For models with discrete state variables, different approaches in large deviation theory are needed \citep{Wainrib:2013bs}.  However, our approach can be adapted to work in systems that deviate from several of the other characteristics. For instance, characteristic~1 is not a limitation of the quasi-potential framework; \citet{Kifer:1990df} describes how analogous concepts can be applied to discrete-time Markov chains \citep{Kifer:1990df,Faure:2014eu}. Variable transformations (see appendix~\ref{subsec:ONS}) can be used to compute quasi-potentials for systems that deviate from characteristic~3 (e.g.~those with noise terms of unequal intensity ($\sigma_{1}\neq\sigma{2}$), noise that scales with population density (demographic stochasticity; $\sigma_{i}\,\sqrt{X_{i}}\,dW_{i}$), or multiplicative environmental stochasticity ($\sigma_{i}\,X_{i}\,dW_{i}$) \citep{Hakoyama:2000jz}). Perturbations to parameters rather than state variables can be accommodated by explicitly modeling the parameter as a state variable with its own differential equation \citep{Allen:2007ww}. A similar approach can be applied to models with colored noise (i.e., models that do not have characteristic~5). The noise process itself can be explicitly modeled as a state variable with its own differential equation (e.g., an Ornstein-Uhlenbeck process). Unfortunately, increasing the dimensionality of the state space in these ways makes the process of numerically calculating the quasi-potential even more challenging. Given the pace of development of numerical techniques \citep{Cameron:2012ex}, however, it is conceivable that solving such systems will soon be more practical.

Characteristic~6, which states that noise occurs continually with low intensity, is central to the quasi-potential framework. The expressions relating the quasi-potential to steady-state probability distributions and mean first passage times are based on the assumption that the noise intensity is very small. As a rule of thumb, these approximations are only useful when $\sigma^{2}$ is much less than $2\,\Delta V_{0}$, where $\Delta V_{0}$ is the difference in the quasi-potential between the stable equilibrium and the saddle. In appendix~\ref{subsec:MFPTa}, we provide details on how mean first passage time scales with noise intensity, and present a numerical examination of these concepts applied to example 2. For systems that experience extreme events and external shocks (e.g., natural disasters, extreme climactic conditions, invasive species introductions, etc.), the quasi-potential no longer provides complete information. If a shock directly impacts the state variable (e.g., if the lake system in example 1 were to receive a massive pulse of phosphorous run-off), the ball in the ball-in-cup diagram would experience a large, instantaneous horizontal displacement (perhaps skipping over intervening valleys and hills). If the system reverts to deterministic dynamics, or stochastic dynamics with lower\-intensity perturbations after the shock, the quasi-potential will still be useful for describing the system's response after the shock. In the presence of large shocks, though, the quasi-potential loses its ability to make probabilistic predictions. If a shock impacts the state variable indirectly (e.g., if an invasive species entered the lake and fundamentally altered the phosphorous cycling), the shape of the quasi-potential surface would change dramatically. The interaction between a dynamically changing quasi\-potential surface and state\-variable noise would be difficult to analyze using the methods presented here.

The three examples in this manuscript show that the quasi-potential often provides a more informative stability metric than traditional linear analysis. Linear stability is much easier to measure in the field, though. This can be done by slightly perturbing a system and measuring the time it takes to return to equilibrium. Before the quasi-potential can be calculated, a model must be fit to observed data and validated.  This limitation is also shared by other methods for analyzing systems with alternative stable states, which depend explicitly \citep[e.g.][]{Boettiger:2012jc} or implicitly \citep[e.g.][]{Dakos:2008dy} on underlying models. Fortunately, carefully controlled experiments \citep{Dai:2012gx} and advances in model\-fitting \citep{Ives:2008kj} point toward a promising future for the empirical study of shifts between alternative stable states through models.  
\medskip
\section*{A Path Through the Quagmire of Stability Concepts}

Systems with alternative stable states are only interesting when perturbations
can cause shifts between states; when these stochastic perturbations are continual and random, as in most ecological systems, stochastic models are appropriate. When state and time
variables are continuous, stochastic differential equations like \eqref{gradient3} are the
best option. The
three examples presented in this paper show that the quasi\-potential
provides a useful way to study such stochastic differential equation
models. In particular, it provides a way to quantify the relative
stability of alternative stable states.

Unfortunately, many notions of stability were developed for a deterministic
context, and these can be misleading when applied to stochastic systems
(as in examples 1 and 2). Our goal is not
to add to the existing tangle of stability definitions \citep{Grimm:1997tg}, but rather to provide
a clarifying mathematical interpretation. Many existing definitions
can be related to the ball\-in\-cup heuristic, and the quasi\-potential
shows that this metaphor has a useful and rigorous mathematical meaning.
The translation between mathematical model and potential surface is
easy in gradient systems (in particular, for one\-dimensional systems,
which are always gradient systems). The translation for more general
systems is less obvious, but the quasi\-potential fills that need.

\begin{figure}[ht]
\centering
\begin{subfigure}[b]{0.7\textwidth}
\caption{}
\includegraphics[width=\textwidth]{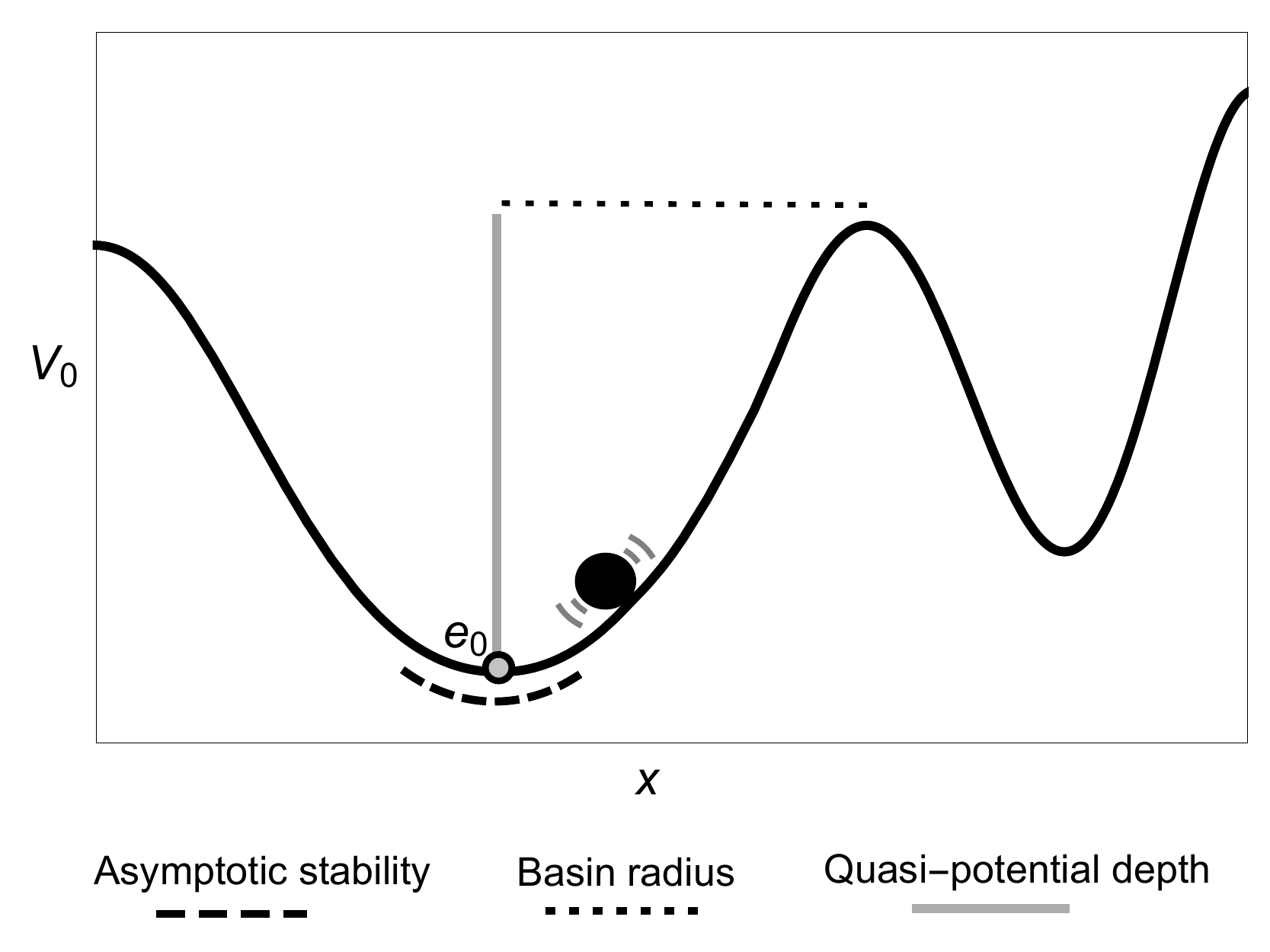}
\end{subfigure}
\vspace*{3mm}
\begin{subfigure}[b]{1.0\textwidth}
\caption{}
\includegraphics[width=\textwidth]{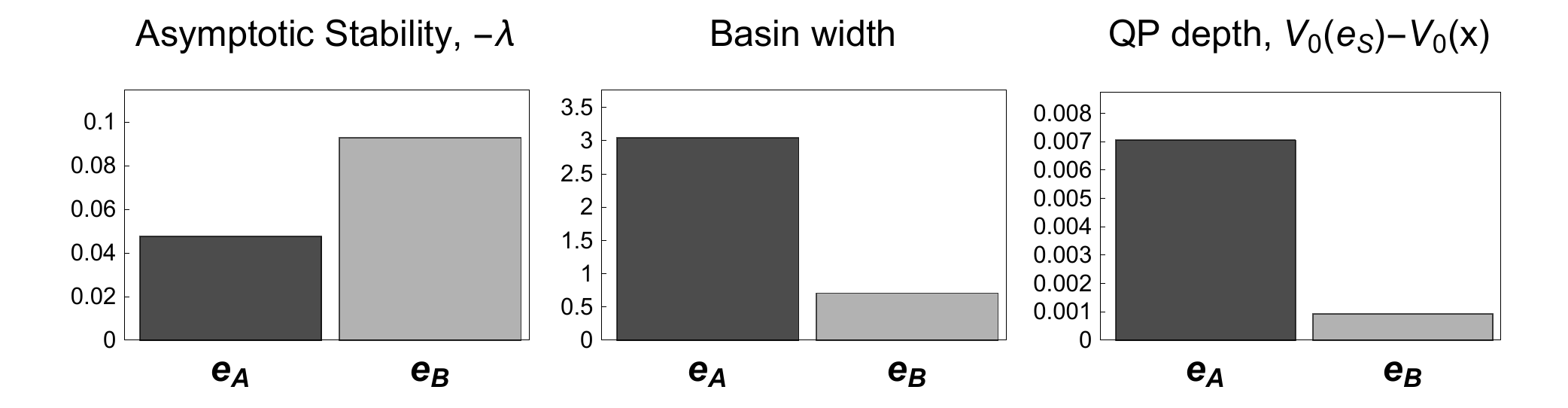}
\end{subfigure}
\caption{\small {\label{Fig6} \textbf{(a)} A schematic diagram of the relationship between various concepts of stability, as related to the quasi-potential and $V_{0}$. \textbf{(b)} A comparison of three different metrics of stability for the system~\eqref{conres}.}}
\end{figure}

Figure~\ref{Fig6}a is a ball\-in\-cup diagram of the potential for a one\-dimensional system that helps to illustrate several important concepts associated with stability. These concepts are equally relevant for higher dimensional systems, where the ball rolls on a multi\-dimensional surface specified by $V_{0}$ (half the Freidlin\-Wentzell quasi\-potential) instead of a curve.

One metric of stability for an equilibrium $\mathbf{e_{0}}$ is the curvature of $V_0$ at $\mathbf{e_{0}}$ (dashed black line in figure~\ref{Fig6}a). The greater the curvature, the more difficult it is to perturb the system away from $\mathbf{e_{0}}$, and in this sense, the more stable $\mathbf{e_{0}}$ is. In one dimension, the curvature at $\mathbf{e_{0}}$ is $V''(\mathbf{e_{0}})$, which is minus the eigenvalue obtained in linear stability analysis. In higher dimensions, the eigenvalues are again directly related to curvature, now along different planar sections of $V_0$ (see appendix~\ref{subsec:curve}). Thus, measuring the curvature of $V_{0}$ at $\mathbf{e_{0}}$ is equivalent to determining asymptotic stability through linear stability analysis.

Asymptotic stability has a long history in ecology \citep{May:1973wc}. The primary problem with this metric is
that it is purely local -- once a trajectory is perturbed outside of
a tiny neighborhood of an equilibrium, nonlinear effects can come
into play and the approximation is no longer informative. Furthermore, this approach views perturbations as being
isolated one\-time events. With this view, a system is displaced, and
then the dynamics proceed deterministically without further perturbation. In reality,
perturbations often take place on a continual basis. Indeed, as noted
by \citet{Ives:1995tm}, ``To apply generally to ecological communities,
stability needs to be defined for stochastic systems in which environmental
perturbations are continuous and equilibrium densities are never achieved.'' Likewise, \citet{Neubert:1997wk} write, ``real ecosystems are seldom
if ever subject to single, temporally isolated perturbations. Nevertheless,
our analyses, together with most theoretical and experimental studies
of resilience, ignore the effects of continual stochastic disturbances
in the hope that the deterministic results will shed light on the
stochastic case."

A second metric of stability of an equilibrium $\mathbf{e_{0}}$ is the minimum distance between $\mathbf{e_{0}}$ and the boundary of its domain of attraction (dotted line in figure~\ref{Fig6}a). The width of the basin of attraction measures the magnitude of perturbation that a system can sustain and still be guaranteed to return to $\mathbf{e_{0}}$. One problem with this metric
is that, like asymptotic stability, it views perturbations as singular,
isolated events. For this metric, it is only the
boundary of basins of attraction that matter, not the shape or height
of $V_{0}$. If perturbations happen continuously,
the shape and height of the $V_{0}$ are important. Nonetheless, this basin width metric can be extremely useful.

A third metric of stability is the height of $V_{0}$ (gray line in figure~\ref{Fig6}a). \citet{Holling:1973wh} anticipated this concept, and called it resilience, which he explained with ball\-in\-cup diagrams. He defines one aspect of
resilience, writing: ``the height of the lowest point of the basin
of attraction ... will be a measure of how much the forces have to
be changed before all trajectories move to extinction of one or more
of the state variables''. Holling had no way of defining the surface,
and so could not actually quantify notions like {}``height''; the
quasi\-potential solves this problem. Holling's identification of the
difference between asymptotic stability and this definition of resilience (basin height) is hugely important, and it has major consequences for the analysis of alternative stable states.

This third metric is perhaps the most useful of the three we have explored. Unlike the first two metrics, it is
appropriate for use in systems that undergo continuous stochastic perturbations. As we saw in the examples in this paper, it can be used to compute mean first passage times, and is directly related to steady\-state probability densities.

These three metrics of stability can yield conflicting information about alternative stable states. Figure~\ref{Fig6}b shows these three metrics for the equilibria $\mathbf{e}_{A}$ and $\mathbf{e}_{B}$ from example 2. Note that the basin width metric and the quasi\-potential metric show that $\mathbf{e}_{A}$ is more stable than $\mathbf{e}_{B}$, but the asymptotic stability metric shows the reverse. Appendix~\ref{subsec:AnotherEx} demonstrates that the equilibria in a multi-stable system can exhibit any combination of the three stability metrics. That is, one equilibria can be classified as most stable according to the first metric, but not the second or third; or by the first and second, but not the third; etc.

Resilience is a concept closely related to stability, and like stability, it is defined in different ways by different authors. In a large review of the ecological literature, \citet{MyersSmith:2012dr} found that resilience was used in many ambiguous
and contradictory ways. Some authors, like \citet{Holling:1973wh} view stability and resilience as distinct properties; others, like \citet{Harrison:1979uya} define resilience as a single aspect of stability. \citet{Pimm:1984tu} and \citet{Neubert:1997wk} define resilience as essentially the asymptotic stability metric, while \citet{Harrison:1979uya}, \citet{Peterson:1998cn}, and \citet{Gunderson:2000ja} define it as essentially the basin width metric. \citet{Ives:2007jba} defines Holling's resilience using the dominant eigenvalue of the saddle that separates alternative stable states; like the asymptotic stability metric, this is the result of applying a local analysis to the deterministic skeleton of a system.

\citet{Hodgson:2015dm} argue that resilience cannot be quantified by a single metric, and use a potential function to illustrate the different components of resilience, which include latitude (the width of the basin of attraction) and elasticity (the asymptotic stability metric). The quasi-potential framework aids this clarification about resilience by extending it to multi-dimensional systems.

The quasi-potential is also useful for understanding several other concepts related to stability. Reactivity \citep{Neubert:1997wk}
differs from asymptotic stability, in that it quantifies the immediate
(as opposed to long\-term) growth or decay of perturbations. In the
quasi\-potential framework, reactivity is related to the circulatory
component of the vector field. In the neighborhood of asymptotically
stable equilibria with high reactivity, the circulatory component
of the vector field will carry trajectories away from the equilibrium
before bringing them back. 

\citet{Harrison:1979uya} defined resistance
as the ability of a system to avoid displacement during a time of
stress. The stress is quantified in terms of an environmental parameter
distinct from the state variables, and hence the interpretation of
resistance depends on the parameter under examination. Resistance is best viewed as a measure of how
dramatically $V_{0}$ changes due to environmental parameter changes.

Finally, Harrison defined persistence as the ability of a system to stay in
a given range when continual perturbations are applied. He notes that
this is the property that is most biologically useful, and that stochastic
differential equations are the best mathematical modeling tool to
assess it. Unlike his definitions of resilience and resistance, this
definition views the dynamics of the system as stochastic and subject
to continual perturbations. He was unable to venture
far with the mathematical analysis for this definition, but the quasi\-potential provides a way forward. Mathematically, persistence can be defined as the first passage time for a system to leave a specified domain, which is directly related to the quasi\-potential. Thus Harrison's persistence is another manifestation of the quasi\-potential.

Despite the confusing array of stability concepts currently used in ecology, we believe that the quasi\-potential concept provides hope for clarity. The three metrics associated with the quasi\-potential show how many of these concepts are deeply related (figure~\ref{Fig6}).
The mathematics developed
by Freidlin and Wentzell \citeyearpar{Freidlin:2012wd}, coupled with numerical advances by Cameron \citeyearpar{Cameron:2012ex},
make the quasi\-potential a practical and accessible tool for ecologists
to study alternative stable states. This paper's goal is to demonstrate
the utility of the quasi\-potential, and to properly position it in
terms of existing ecological ideas.
\medskip
\section*{Acknowledgements}
This work was supported by a Complex Systems Scholar grant to K.C.A.~from the James S.~McDonnell Foundation. Special thanks to M.K.~Cameron for assistance with implementing the quasi\-potential analysis and for providing C code. C.~Boettiger and an anonymous reviewer provided valuable feedback that improved the quality of this paper. We thank S.~Catella, K.~Dixon, C.~Moore, C.~Stieha, A.~Barbaro, A.~Alsenafi, R.~Snyder, J.~Burns, and the rest of the CWRU ecology group for helpful discussions on earlier versions of this manuscript.
\medskip
\bibliography{QPStabKCA}
\bibliographystyle{amnatnat}

\newpage

\newpage

\newpage

\newpage

\newpage

\FloatBarrier

\newpage
\appendix
\section*{Appendices}
\renewcommand{\thesubsection}{\Alph{subsection}}
\medskip
\subsection{Stochastic Differential Equations}
\label{subsec:SDEAppendix}

In example 1, we briefly described a stochastic differential equation model for lake eutrophication. In this section of the appendix, we provide background information about stochastic differential equations. A random variable $X$ is a variable whose value is subject to chance. When a specific outcome $X=x$ is observed, it is called a realization. A stochastic process $X(t)$ is a family of random variables indexed by the parameter $t$, which usually represents time. Time can be measured discretely or continuously; this latter case falls in the realm of stochastic differential equations. A realization, $X(t)=x(t)$, is obtained when the stochastic process is observed at each time $t$. Note that a realization $x(t)$ is a deterministic function of time.

A continuous\-time stochastic process of particular importance is the Wiener process, also known as Brownian motion, and denoted by $W(t)$. This process can be visualized as the limit of a discrete time random walk, which changes by an amount $\Delta W$  per each time step $\Delta t$. Each increment $\Delta W$ is selected from a normal distribution with mean $0$ and variance $\Delta t$. The Wiener process is the limit of this random walk as $\Delta t\rightarrow0$. It turns out that the Wiener process is completely characterized by three properties:
\begin{myenumerate}
\item $W\left(0\right)=0$
\item $W\left(t\right)$ is almost surely continuous everywhere. This means
that, with 100\% probability, a realization will be continuous (aside
from possibly a few bad points, which have measure zero).
\item If $0\leq s_{1}<t_{1}\leq s_{2}<t_{2}$, then $W(t_{1})-W(s_{1})$
is normally distributed with mean zero and variance $t_{1}-s_{1}$,
$W(t_{2})-W(s_{2})$ is normally distributed with mean zero and variance
$t_{2}-s_{2}$, and $W(t_{1})-W(s_{1})$ and $W(t_{2})-W(s_{2})$
are independent.
\end{myenumerate}
The Reimann\-Stieljes integrals of elementary calculus are defined
as the limits of finite sums. Integration with respect to a Wiener
process can be defined in a similar way. The It\^{o} integral of the function
$h$ of a stochastic process $X(t)$ over the interval $[0,T]$ is
defined as:
\begin{equation}
\int_{0}^{T}h(X(t))\, dW=\lim_{n\rightarrow\infty}\sum_{i=0}^{n-1}h(X(t_{i}))\,(W(t_{i+1})-W(t_{i})),
\end{equation}
where $\left\{ [t_{i},t_{i+1})\right\} _{i=0}^{n}$ is a partition
of $[0,T]$. Note that this integral is a stochastic process itself; each realization of $X$ and $W$ leads to a different realization
of the integral. In the It\^{o} integral, $h(X(t))$ is evaluated at the
left end points of the intervals of the partition. If a trapezoidal
rule is used instead, then the result is the Stratonovich integral.
In this paper, we use the It\^{o} integral, because of the way it discriminates
between the past and the future. A process $X(t)$ is called ``non-anticipating"
if its value at $t$ is independent of values of $W(s)$, for $s>t$.
If $X(t)$ is non\-anticipating, then the It\^{o} integral defined above
is, too. The Stratonovich integral is not, because calculating the
integral at time $s$, $t_{i}\leq s<t_{i+1}$ requires knowledge of
$X(t_{i+1})$. Basically, the It\^{o} integral cannot {}``see into the
future'', while the Stratonovich integral can.

Having defined integration with respect to a Wiener process, we can
now define a stochastic differential equation. Consider a deterministic
autonomous differential equation, 
\begin{equation}
\frac{dx}{dt}=f(x).
\end{equation}
In a small time period $\Delta t$, the variable $x$ changes by an
amount of approximately $\Delta x=f(x)\,\Delta t$. Now suppose that
the variable $x(t)$ is subject to random disturbances, and hence
is a stochastic process $X(t)$. To approximate the value of this
stochastic process at time $T$, we discretize time into $m$ small
intervals, each of length $\Delta t$. Let $X_{i}=X\left(i\Delta t\right)$,
and $\Delta X_{i}=X_{i+1}-X_{i}$. During a time period of length
$\Delta t$, there are probably many small perturbations that affect
$X$; if they have finite variance, then by the central limit theorem,
adding these small perturbations up yields a normally distributed
random variable. We will assume that this accumulated perturbation
over a time period of length $\Delta t$ has mean 0 and variance $\sigma^{2}\Delta t$
(the linear relationship with $\Delta t$ is required in order for
$X(T)$ to have finite, non\-zero variance in the continuous time limit).
Therefore, the change in the stochastic process over a time interval
of length $\Delta t$ can be written as:
\begin{equation}
\Delta X_{i}=f(X_{i})\Delta t+\sigma\Delta W_{i},
\end{equation}
where $\Delta W_{i}$ is normally distributed with mean 0 and variance
$\Delta t.$ Adding up the changes in the process over the time interval
$[0,T]$ yields 
\begin{equation}
X(T)=X(0)+\sum_{i=1}^{m}f(X_{i})\Delta t+\sigma\sum_{i=1}^{m}\Delta W_{i},
\end{equation}
which suggests an integral equation for the continuous time limit,
\begin{equation}
\label{inteq}
X(T)=X(0)+\int_{0}^{T}f(X)dt+\sigma\int_{0}^{T}dW.
\end{equation}
If the intensity of perturbations depend on the value of $X$, then equation~\eqref{inteq} can be generalized to 
\begin{equation}
\label{int}
X(T)=X(0)+\int_{0}^{T}f(X)dt+\sigma\int_{0}^{T}g\left(X\right)dW
\end{equation}
The integrals in equation~\eqref{int} make the notation cumbersome. In light of this,
a modified notation is used. The stochastic differential equation
\begin{equation}
\label{stochdiff}
dX=f(X)\, dt+\sigma\, g(X)\, dW
\end{equation}
formally means that $X(t)$ is a solution to equation~\eqref{stochdiff}. Note that $\frac{dW}{dt}$
does not exist, because the sample paths of $W(t)$ are almost surely
nowhere differentiable. This is why the notation in equation~\eqref{stochdiff} is used; it reminds
us that $X(t)$ is defined by the integral equation~\eqref{int}.
\medskip

\subsection{Freidlin-Wentzell quasi-potential}
\label{subsec:FWAppendix}
In this section, we provide a more formal definition of the Freidlin\-Wentzell quasi\-potential. Consider a system of stochastic differential equations
\begin{equation}
\label{gradient}
d\mathbf{X}=f(\mathbf{X})\,dt+\sigma\,g(\mathbf{X})\,d\mathbf{W},
\end{equation}
where $\mathbf{X}=\left(X_{1},\,\ldots,\, X_{n}\right)$ is a vector of
state variables, $\mathbf{W}=\left(W_{1},\,\ldots,\, W_{m}\right)$ is
a vector of $m$ independent Wiener processes. Vectors in this paper should be interpreted as column vectors. The lower\-case notation $\mathbf{x}=\left(x_{1},\,\ldots,x_{n}\right)$ is used to indicate a point (as opposed to a stochastic process). $f$
is a vector field that is the deterministic skeleton of the system. $g(\mathbf{x})$ is a matrix that determines how the different noise sources affect the state variables, and $\sigma$ is the noise intensity. For simplicity, we will focus on the case where $m=n$ and $g(\mathbf{x})$ is the identity matrix, which represents constant\-intensity isotropic noise, affecting each state variable with equal intensity. Under these assumptions, equation~\eqref{gradient}
can be written as
\begin{equation}
\label{gradient2}
d\mathbf{X}=f(\mathbf{X})\,dt+\sigma\,d\mathbf{W}.
\end{equation}
In appendix~\ref{subsec:ONS}, we will return to the general case~\eqref{gradient}, but constant, isotropic
noise provides a useful starting point.
If there exists a function
$U(\mathbf{x})$ such that $f=-\nabla U$,
then the differential equations are called a gradient system, and the function $U$ is called
a potential function. Like one\-dimensional systems, a multi\-dimensional gradient system
can be viewed with the ball\-in\-cup framework. For $n=2$, the relevant
metaphor is a ball rolling on a two\-dimensional surface specified
by the function $U(\mathbf{x})$. For $n\geq3$,
the situation is difficult to visualize, but the same general intuitive
aspects hold. The steady\-state
probability distribution of higher\-dimensional gradient systems is related to the potential $U$ in the same way as in~\eqref{steadystate}, except $\mathbf{x}$ replaces $x$ and $Z$ is obtained from an $n$-dimensional integral.
Expressions for the mean first passage time between stable equilibria
separated by a saddle are similar to the one\-dimensional case as well.

Unfortunately, gradient systems are a very special situation. In most
cases of~\eqref{gradient2}, there will not exist a function $U$ satisfying $f=-\nabla U$.
For these non\-gradient systems, we cannot use a potential function
to quantify stability, as we did in example 1. In what follows, we develop an approach that is conceptually analogous but applicable to non\-gradient systems.

In the following, we will use the concept of logarithmic equivalence, denoted by $\asymp$. We write $f(x)\asymp e^{\kappa h(x)}$ if
\begin{equation}
\lim_{\kappa\rightarrow\infty}\kappa^{-1}\ln(f(x))=h(x).
\end{equation}
The Freidlin\-Wentzell approach is to obtain a large deviation principle for trajectories $x(t)$ of~\eqref{gradient2}. In this context, a large deviation principle is an asymptotic rule that determines how likely it is for realizations of~\eqref{gradient2} to depart from a given path. To make this concrete, let $\mathbf{a}$ be an asymptotically stable equilibrium of $\mathbf{f}$ in~\eqref{gradient2}. Let $\mathbf{b}\in\mathbb{R}^{n}$ and $T>0$. Let $\Theta_{T}$ be the set of all absolutely continuous paths $\theta:[0,T]\rightarrow\mathbb{R}^{n}$
such that $\theta(0)=\mathbf{a}$ and $\theta(T)=\mathbf{b}$. We will study the probability that a realization $\mathbf{x}_{\sigma}(t)$ of~\eqref{gradient2} with noise intensity $\sigma$ and with $\mathbf{x}_{\sigma}(0)=\mathbf{a}$ and $\mathbf{x}_{\sigma}(T)=\mathbf{b}$ stays close to $\theta\in\Theta_{T}$. A large deviation principle declares that there exists a $\delta_{0}>0$ such that, if $0<\delta<\delta_{0}$, then
\begin{equation}
\label{largedev}
\Pr\left\{ \sup_{0\leq s\leq T}\left|\mathbf{x}_{\sigma}(s)-\theta(s)\right|<\delta\right\} \asymp\exp\left(-\frac{S_{T}(\theta)}{\sigma^{2}}\right),
\end{equation}
where the logarithmic equivalence holds as $\sigma\rightarrow0$.
The functional $S_{T}:\Theta_{T}\rightarrow[0,\infty)$ is called the action, and it is defined by 
\begin{equation}
S_{T}(\theta)=\frac{1}{2}\int_{0}^{T}\left|f(\theta(t))-\dot{\theta}(t)\right|^{2}dt.
\end{equation}
Note that $S_{T}$ measures how much $\dot{\theta}$ deviates from the vector
field $f$. If $S_{T}(\theta)=0$, then $\theta$ is a
trajectory of the deterministic system, $\frac{dx}{dt}=f(x)$. The action $S_{T}$ is related to the probability distribution of $\mathbf{X}$ by
\begin{equation}
\label{largedev2}
\lim_{\sigma\rightarrow0}\sigma^{2}\ln\left(\Pr\left\{\mathbf{X}(T)\in\Omega\vert \mathbf{X}(0)=\mathbf{a}\right\} \right)=-\inf_{\theta\in\Theta_{T}}\left\{ S_{T}(\theta)\vert\theta(0)=\mathbf{a},\theta(T)\in\Omega\right\},
\end{equation}
where $\Omega$ is a domain in $\mathbb{R}^{n}$. For details on the technical assumptions behind this relationship, see \citet{Freidlin:2012wd}.
To get from $\mathbf{a}$ to $\mathbf{b}$ in a {}``likely'' way, the action should
be made as small as possible. This motivates the definition of the Freidlin\-Wentzell quasi\-potential (or simply quasi\-potential), $\Phi_{\mathbf{a}}:\mathbb{R}^{n}\rightarrow[0,\infty)$,
\begin{equation}
\Phi_{\mathbf{a}}(\mathbf{b})=\inf_{T>0,\theta\in\Theta_{T}}\left\{ S_{T}(\theta)\vert\theta(0)=\mathbf{a},\theta(T)=\mathbf{b}\right\} .
\end{equation}
Note that the infimum is taken over paths of all durations (that is, all times $T>0$).

The quasi\-potential is the value of the action for the minimum\-action path (i.e., the most likely path) between $\mathbf{a}$ and $\mathbf{b}$. It is closely
related to first passage times from domains of attraction. If $D$ is a region contained within the domain of attraction of $\mathbf{a}$, then the expected time until a trajectory exits $D$, $\tau_{\mathbf{a}}^{\partial D}$, is given by
\begin{equation}
\lim_{\sigma\rightarrow0}\sigma^{2}\ln(\tau_{\mathbf{a}}^{\partial D})=\inf_{\mathbf{x}\in\partial D}\Phi_{\mathbf{a}}(\mathbf{x}).
\end{equation}

The quasi\-potential need not be defined solely in terms of an isolated asymptotically stable equilibrium $\mathbf{a}$. \citet{Cameron:2012ex} generalized the quasi\-potential, and defined it for compact sets. This generalization allows the quasi\-potential to be determined for limit cycles (as demonstrated in example 3). A different approach to generalizing the quasi-potential to compact sets can be can be found in \citet{Freidlin:2012wd}. Cameron's generalization requires considering the geometric action \citep{Heymann2008, Heymann2008b}, which we will denote by $S^{*}$. Suppose that $\theta\in\Theta_{T}$, and $\psi(\nu)$ is a reparameterization of $\theta$ such that $\psi(0)=\theta(0)$ and $\psi(\nu_{0})=\theta(T)$. Then the geometric action is
\begin{equation}
S^{*}(\psi)=\int_{0}^{\nu_{0}}\left|f(\psi(\nu))\right||\dot{\psi}(\nu)|-f(\psi(\nu))\cdot\dot{\psi}(\nu)\, d\nu.
\end{equation}
The value of $S^{*}$ is independent of the parameterization of $\psi$. If $A$ and $B$ are compact sets in $\mathbb{R}^n$, then the quasi\-potential can be defined by
\begin{equation}
\Phi_{A}(B)=\inf\left\{ S^{*}(\psi)\vert\psi(0)\in A,\psi(\nu_{0})\in B\right\}.
\end{equation}
\medskip

\subsection{A global quasi\-potential}
\label{subsec:GQP}
In systems with multiple stable equilibria, it is desirable to obtain a global quasi\-potential that describes how trajectories switch between states. In the preceding section, the quasi\-potential was defined in terms of a stable equilibrium $\mathbf{a}$. Suppose now that there are two stable equilibria, $\mathbf{a}_{1}$ and $\mathbf{a}_{2}$, with corresponding domains of attraction $D_{1}$ and $D_{2}$. The action functionals can be used to obtain $\Phi_{\mathbf{a}_{1}}$ and $\Phi_{\mathbf{a}_{2}}$, but these quasi\-potentials are of limited utility outside of $D_{1}$ and $D_{2}$, respectively. The minimum action path from $\mathbf{a}_{1}$ to $\mathbf{a}_{2}$ will follow streamlines of the vector field once it enters $D_{2}$. This will result in no accumulated work; hence $\Phi_{\mathbf{a}_{1}}$ will be flat along streamlines in $D_{2}$. The quasi\-potentials both describe dynamics well within their domains of attraction, but in order to create a complete surface in the spirit of a classical potential function, it is necessary to combine the two. This is easily accomplished if there is a single saddle point $\mathbf{s}$ that lies on the separatrix between $D_{1}$ and $D_{2}$. We find the constant
\begin{equation}
C = \Phi_{\mathbf{a}_{1}}(\mathbf{s})- \Phi_{\mathbf{a}_{2}}(\mathbf{s})
\end{equation}
so that $\Phi_{\mathbf{a}_{2}}^{*} = \Phi_{\mathbf{a}_{2}} + C$ agrees with $\Phi_{\mathbf{a}_{1}}$ at $\mathbf{s}$. Finally, we compute the global quasi\-potential $\Phi$ as
\begin{equation}
\Phi (\mathbf{x}) = \min \left( \Phi_{\mathbf{a}_{1}}(\mathbf{x}),\Phi_{\mathbf{a}_{2}}^{*}(\mathbf{x}) \right).
\end{equation}
More complicated cases can arise when domains of attraction are connected by more than one saddle \citep{Freidlin:2012wd}. For details about how to combine local quasi-potentials into a global quasi-potential in these more complicated cases, see \citet{Freidlin:2012wd}, \citet{Moore2015}, and \citet{roy1994}.
\medskip

\subsection{Small noise expansion of $V$}
\label{subsec:SmallNAppendix}
This section describes the relationship between $V$ and $V_{0}$, and shows the derivation of the Hamilton\-Jacobi equation for $V_{0}$. The Fokker\-Planck equation associated with the two\-dimensional version of \eqref{gradient2} is \eqref{FP}. Under relatively mild conditions on the function $f$ \citep[for details, see][]{Freidlin:2012wd}, there will exist a steady\-state probability distribution
\begin{equation}
p_{s}(\mathbf{x})=\lim_{t\rightarrow\infty}p(\mathbf{x},t).
\end{equation}

Steady\-state distributions can often be approximated by very long\-time realizations. Determining when such an approximation holds is the subject of ergodic theory \citep[see][]{arnold2010random}. Approximations to steady\-state distributions for the consumer\-resource in example 2 (i.e., the system \eqref{conres}) are shown in figure~S1. Each panel corresponds to a different noise intensity, $\sigma$. Qualitatively, figure~S1 confirms that trajectories spend more time near $\mathbf{e}_{A}$ than $\mathbf{e}_{B}$,
and hence a sensible stability metric should classify $\mathbf{e}_{A}$ as more stable
than $\mathbf{e}_{B}$. However, it also clearly shows that the steady\-state distribution depends on the noise level; each choice of $\sigma$ yields a different distribution. If one is interested in the general properties of the system, and not just the steady\-state distribution for a specific noise intensity, then steady\-states distributions are of limited utility.

The ``effective potential" (not to be confused with the potential or quasi\-potential) is defined as
\begin{equation}
\label{effpot}
V(\mathbf{x})=-\frac{\sigma^{2}}{2}\ln{p_{s}(\mathbf{x})}+C,
\end{equation}
where $C$ is a constant. The effective potential's relationship with the steady\-state distribution makes it a helpful tool. The peaks of the steady\-state distribution correspond to valleys of the effective potential, and vice versa. There are two reasons why we do not adopt the effective potential as a stability metric in this paper. First, the effective potential depends on $\sigma$, and hence suffers from the same issue as the steady\-state distribution. In the ball\-in\-cup metaphor, the noise intensity $\sigma$ determines the perturbations of the ball as it rolls, rather than determining the shape of the landscape. Second, the effective potential is not a Lyapunov function for the deterministic system, so a trajectory of a system with zero noise does not necessarily move downhill. Finally, a decomposition based on the gradient of $V$ is not orthogonal. Despite these shortcomings, the effective potential is closely related to the quasi\-potential.
Solving \eqref{effpot} for $p_{s}(\mathbf{x})$ yields
\begin{equation}
p_{s}(\mathbf{x})=e^{\frac{2\,C}{\sigma^{2}}}e^{-\frac{2\,V(\mathbf{x})}{\sigma^{2}}}.
\end{equation}
Substituting this into the Fokker\-Planck equation yields
\begin{equation}
\label{fpresults}
\left|\nabla V\right|^{2}+f_{1}\frac{\partial V}{\partial x_{1}}+f_{2}\frac{\partial V}{\partial x_{2}}-\frac{\sigma^{2}}{2}\left(\nabla^{2}V+\frac{\partial f_{1}}{\partial x_{1}}+\frac{\partial f_{2}}{\partial x_{2}}\right)=0.
\end{equation}
To simplify this equation, we consider how the system behaves for small noise values, and expand $V$ in terms of the small parameter $\epsilon=\frac{\sigma^{2}}{2}$. This yields
\begin{equation}
V(\mathbf{x})=\sum_{i=0}^{\infty}V_{i}(\mathbf{x})\epsilon^{i},
\end{equation}
where $V_{i}$ is the coefficient function associated with order $\epsilon^{i}$. Inserting this into \eqref{fpresults} and retaining lowest-order terms, we obtain the Hamilton\-Jacobi equation for $V_{0}$
\begin{equation}
\left|\nabla V_{0}\right|^{2}+f_{1}\frac{\partial V_{0}}{\partial x_{1}}+f_{2}\frac{\partial V_{0}}{\partial x_{2}}=0.
\end{equation}
\medskip

\subsection{Hamilton\-Jacobi equation for the quasi\-potential}
\label{subsec:HJEQP}
By deriving the Hamilton\-Jacobi equation for the quasi\-potential, we can verify the relationship $\Phi=2V_{0}$. This relationship is crucial. We described key properties about $V_{0}$ concerning the effective potential, the steady\-state probability distribution, the Lyapunov property, and the orthogonality of the decomposition $f=-\nabla V +Q$; in appendix B, we described properties of $\Phi$. The relationship $\Phi=2\,V_{0}$ shows that these functions share those properties; they only differ by multiplication of a scalar.

Bellman's Principle from optimal control theory can be used to derive the Hamilton\-Jacobi equation for $\Phi$. We sketch the proof from \citet{Cameron:2012ex}. The calculation of $\Phi_{A}(\mathbf{x})$, the value of the quasi\-potential starting at a compact set $A$ and going to a point $\mathbf{x}$, can be viewed as an optimal control problem. We seek to minimize the value function $\Phi_{A}(\mathbf{x})$ by choosing an optimal path $\psi(\nu)$. This path is controlled by the velocity vector $\dot{\psi}(\nu)$. We are free to choose the parameterization of $\psi(\nu)$, so we select one where the velocity vector has unit magnitude at every point. The optimal control problem amounts to determining the tangent direction $\dot{\psi}(\nu)$ for each $\nu$, so that the resulting path minimizes the action. Bellman's Principle essentially turns this problem into a recursive equation. Heuristically, one can imagine the last segment of an optimal path $\psi(\nu)$ from $\psi(0)\in A$ to $\psi(K)=\mathbf{x}$. This last segment is specified by the parameter values $\nu\in[K-\delta,K]$. Clearly this optimal path will be optimal over the interval $[K-\delta,K]$. Therefore, if one knows the optimal path up to parameter value $K-\delta$, one knows the remainder of the path as well. Mathematically, this principle takes the form:
\begin{equation}
\Phi_{A}(\mathbf{x})=\inf_{\dot{\psi}\in\mathbb{S}^{n-1}}\left\{ \int_{K-\delta}^{K}\left|f(\psi(\nu))\right|-f(\psi(\nu))\cdot\dot{\psi}(\nu)\, d\nu+\Phi_{A}\left(\psi(K-\delta\right))\right\}.
\end{equation}
A small $\delta$ expansion yields:
\begin{equation}
\Phi_{A}(\mathbf{x})=\inf_{\dot{\psi}\in\mathbb{S}^{n-1}}\left\{ (\left|f(\mathbf{x})\right|-f(\mathbf{x})\cdot\dot{\psi})\delta+\Phi_{A}(\mathbf{x)}-\nabla \Phi_{A}(\mathbf{x})\cdot\dot{\psi}\,\delta\right\}.
\end{equation}
Solving this equation is equivalent to solving:
\begin{equation}
\label{preHJ}
\inf_{\dot{\psi}\in\mathbb{S}^{n-1}}\left\{ \left|f(\mathbf{x})\right|-f(\mathbf{x})\cdot\dot{\psi}-\nabla \Phi_{A}(\mathbf{x})\cdot\dot{\psi}\right\} =0.
\end{equation}
Using the Cauchy\-Schwarz inequality, one finds that the infimum of the left\-hand side of~\eqref{preHJ} occurs when
\begin{equation}
\dot{\psi}=\frac{f(\mathbf{x})+\nabla \Phi_{A}(\mathbf{x})}{|f(\mathbf{x})|}.
\end{equation}
Substituting this into \eqref{preHJ} yields
\begin{equation}
\left|\nabla \Phi_{A}\right|^{2}+2\,\nabla \Phi_{A}\cdot f=0.
\end{equation}
Comparing this to \eqref{HJE}, we can see that, if solutions exist, they have the relationship $\Phi_{A}=2\,V_{0}$. Classical solutions do not always exist for the Hamilton\-Jacobi equation, so it is often necessary to consider a class of weak solutions called ``viscosity solutions" \citep{Sethian:2001wx, Crandall:1983vt, crandall1984}. When a classical solution does exist, it coincides with the viscosity solution.
\medskip

\subsection{Other noise structures}
\label{subsec:ONS}
This paper focuses on the case $g(\mathbf{x})=I$ in equation~\eqref{gradient}, where $I$ is the identity matrix. The quasi\-potential can be calculated for more general cases. Such a generalization
requires a modification in the definition of the action:
\begin{equation}
S_{T}(\theta)=\frac{1}{2}\int_{0}^{T}\sum_{i,j}q_{i,j}(\theta(t))\left(f_{i}(\theta(t))-\dot{\theta_{i}}(t)\right)\left(f_{j}(\theta(t))-\dot{\theta_{j}}(t)\right)dt.
\end{equation}
where $q(x)=\left(g(x)g^{T}(x)\right)^{-1}.$ The large deviation relationships,~\eqref{largedev} and~\eqref{largedev2}, are still valid. The Hamilton\-Jacobi equation for~\eqref{gradient} is
\begin{equation}
\sum_{i,j}q_{i,j}\frac{\partial\Phi}{\partial x_{i}}\frac{\partial\Phi}{\partial x_{j}}+2\,\nabla \Phi\cdot f=0.
\end{equation}
Alternatively, one can find a transform of~\eqref{gradient} that turns the system into the form~\eqref{gradient2}, compute the quasi\-potential in these new coordinates, and then back\-transform to the original coordinates. For the system
\begin{equation}
\begin{gathered}
dX_{1}=f_{1}(X_{1},\,X_{2})\,dt+\sigma\,g_{1}\,dW_{1}\\
dX_{2}=f_{2}(X_{1},\,X_{2})\,dt+\sigma\,g_{2}\,dW_{2},
\end{gathered}
\end{equation}
where $g_{1}$ and $g_{2}$ are constants, the appropriate transform is $\tilde{X}_{1}=g_{1}^{-1}\,X_{1}$, $\tilde{X}_{2}=g_{2}^{-1}\,X_{2}$. For the system
\begin{equation}
\begin{gathered}
dX_{1}=f_{1}(X_{1},\,X_{2})\,dt+\sigma\,g_{1}\,X_{1}\,dW_{1}\\
dX_{2}=f_{2}(X_{1},\,X_{2})\,dt+\sigma\,g_{2}\,X_{2}\,dW_{2},
\end{gathered}
\end{equation}
the appropriate transform is $\tilde{X}_{1}=g_{1}^{-1}\,\ln (X_{1})$, $\tilde{X}_{2}=g_{2}^{-1}\,\ln (X_{2})$.
\medskip

\subsection{Curvature}
\label{subsec:curve}
The concept of curvature is more nuanced for surfaces than it is for curves. The principal curvatures of the surface specified by $V_{0}$ at $\mathbf{e_{0}}$ are the largest and smallest curvatures of the one\-dimensional normal sections at $\mathbf{e_{0}}$. A normal section is obtained by intersecting a plane containing the normal vector of the surface $V_{0}$ at $\mathbf{e_{0}}$ with $V_{0}$. The principal curvatures correspond to the eigenvalues of the Hessian matrix of $V_0$. In the gradient case, $f=-\nabla V_{0}$, so the Hessian matrix of $V_{0}$ is simply the negative of the Jacobian matrix of $f$. In other words, the principal curvatures of the surface $V_{0}$ are the eigenvalues obtained from the linear stability analysis (except with the sign changed).
\medskip

\subsection{Mean first passage time asymptotics}
\label{subsec:MFPTa}
Steady-state probability densities and mean first passage times can be determined from the quasi-potential, but only in the small-noise limit. These quantities are often expressed in terms of logarithmic equivalence as $\sigma\rightarrow0$. Accurate calculation involves not just the exponential part of the relationship, but also the prefactor. For a gradient system with potential $U$, the mean first passage time $\tau$ to transition from a stable equilibrium $\mathbf{x}$ to a saddle $\mathbf{z}$ is \citep{Bovier:2004tx}:
\begin{equation}
\tau=\frac{2\,\pi}{\left|\lambda_{1}\left(\mathbf{z}\right)\right|}\sqrt{\frac{\left|\det\nabla^{2}U(\mathbf{z})\right|}{\det\nabla^{2}U(\mathbf{x})}}\exp\left(\frac{2\left(U(\mathbf{z})-U(\mathbf{x})\right)}{\sigma^{2}}\right)\left(1+\mathcal{O}\left(\sigma\left|\log\left(\sigma\right)\right|\right)\right)
\end{equation}
$\nabla^{2}U(\mathbf{z})$ is the Hessian of the potential at $\mathbf{z}$, and $\nabla^{2}U(\mathbf{x})$ is the Hessian of the potential at $\mathbf{x}$. $\lambda_{1}(\mathbf{z})$ is the negative eigenvalue of the Hessian at the saddle.
\citet{Bouchet:2015tx} obtained a similar expression has for a non-gradient system with quasi-potential $V$. Their estimate for $\tau$ is:
\begin{equation}
\tau=\frac{2\,\pi}{\left|\lambda_{1}\left(\mathbf{z}\right)\right|}\sqrt{\frac{\left|\det\nabla^{2}V(\mathbf{z})\right|}{\det\nabla^{2}V(\mathbf{x})}}\exp\left(\frac{2\left(V(\mathbf{z})-V(\mathbf{x})\right)}{\sigma^{2}}\right)\exp\left(\int_{-\infty}^{\infty}F(\mathbf{\rho}(t))\, dt\right).
\end{equation}
$\lambda_{1}(\mathbf{z})$ is the unstable eigenvalue of the full deterministic skeleton (not just the quasi-potential) at the saddle. $\mathbf{\rho}(t)$ is the least action path from $\mathbf{x}$ to $\mathbf{z}$:
\begin{equation}
\mathbf{\rho}'(t)=\nabla V\left(\mathbf{\rho}(t)\right)+Q\left(\mathbf{\rho}(t)\right)
\end{equation}
$F$ is the divergence of the circulatory component, $F(\mathbf{x})=\nabla\cdot Q(\mathbf{x})$.

For example 2, we can examine how the mean first passage time estimates from the quasi-potential correspond to simulation results (Figure~\ref{Fig30}). For this example, numerical integration along the least action path suggests that $\int_{-\infty}^{\infty}F(\mathbf{\rho}(t))\, dt\approx0$, so we drop this term in the approximation. Note that the quasi-potential approximations closely match the means of the simulated first passage times. Of course, there will always be outliers, and these can be seen in the tails of the distributions of simulated first passage times in figure~\ref{Fig30}.
\medskip

\subsection{Further example}
\label{subsec:AnotherEx}
In the following, we examine a bistable model that illustrates the different ways that the stability metrics described in this paper can classify stable states. The model is:
\begin{equation}
\label{anotherex}
\begin{gathered}
dX=f_{1}(X,Y)\,dt+\sigma\,dW_{1}\\[5pt]
dY=f_{2}(X,Y)\,dt+\sigma\,dW_{1} .
\end{gathered}
\end{equation}
The deterministic skeleton is given by:
\begin{equation}
\begin{gathered}
f_{1}(x,y)=-2\, a\, b_{1}\, x\,\exp\left(-\left(b_{1}\,x^{2}+b_{2}\,y^{2}\right)\right)-2\, d_{1}\,\left(x-c\right)\exp\left(\left(d_{1}\,\left(x-c\right)^{2}+d_{2}\,\left(y-c\right)^{2}\right)\right)\\[5pt]
f_{2}(x,y)=-2\, a\, b_{2}\, x\,\exp\left(-\left(b_{1}\,x^{2}+b_{2}\,y^{2}\right)\right)-2\, d_{2}\,\left(y-c\right)\exp\left(\left(d_{1}\,\left(x-c\right)^{2}+d_{2}\,\left(y-c\right)^{2}\right)\right)\\. \end{gathered}
\end{equation}
This model does not represent any particular ecological process and was instead chosen for its ability to illustrate the range of relationships that are possible between the stability metrics we discuss. This is a gradient system, with potential function
\begin{equation}
U(x,y)=1-a\,\exp\left(-\left(b_{1}\,x^{2}+b_{2}\,y^{2}\right)\right)-\exp\left(-\left(d_{1}\,\left(x-c\right)^{2}+d_{2}\,\left(y-c\right)^{2}\right)\right).
\end{equation}
For all of the parameter values we consider, the system will have two stable states, $\mathbf{e}_{1}$ and $\mathbf{e}_{2}$, separated by a saddle $\mathbf{e}_{s}$. This example will show that each equilibria can be classified as more stable by any combination of the stability metrics. Without loss of generality, the stable-state with larger $x$~-value, $\mathbf{e}_{2}$, will be more stable according to metric 3 (the basin depth metric).

In case 1, the parameter values are $a=0.9,$ $b_{1}=1,$ $b_{2}=1,$ $c=1.2,$ $d_{1}=1.2,$ $d_{2}=1.2.$ $\mathbf{e}_{2}$ is more stable by all three metrics (Figure S3).
\begin{equation}
\begin{gathered}
\text{Metric 1: }\text{Re}\left(\lambda\left(\mathbf{e}_{1}\right)\right)=-1.16769,\quad
\text{Re}\left(\lambda\left(\mathbf{e}_{2}\right)\right)=-1.75728\\
\text{Metric 2: }\|\mathbf{e}_{1}-\mathbf{e}_{s}\|=0.643635,\quad
\|\mathbf{e}_{2}-\mathbf{e}_{s}\|=0.859567\\
\text{Metric 3: }U\left(\mathbf{e}_{s}\right)-U\left(\mathbf{e}_{1}\right)=0.0842552,\quad
U\left(\mathbf{e}_{s}\right)-U\left(\mathbf{e}_{2}\right)=0.204706
\end{gathered}
\end{equation}

In case 2, the parameter values are $a=0.9,$ $b_{1}=2,$ $b_{2}=2,$ $c=1.8,$ $d_{1}=0.8,$ $d_{2}=0.8.$ $\mathbf{e}_{2}$ is more stable by metric 3 (depth) and metric 2 (basin width) but not metric 1 (linear stability) (Figure S4).
\begin{equation}
\begin{gathered}
\text{Metric 1: }\text{Re}\left(\lambda\left(\mathbf{e}_{1}\right)\right)=-3.51331,\quad
\text{Re}\left(\lambda\left(\mathbf{e}_{2}\right)\right)=-1.59979\\
\text{Metric 2: }\|\mathbf{e}_{1}-\mathbf{e}_{s}\|=1.05186,\quad
\|\mathbf{e}_{2}-\mathbf{e}_{s}\|=1.48722\\
\text{Metric 3: }U\left(\mathbf{e}_{s}\right)-U\left(\mathbf{e}_{1}\right)=0.639468,\quad
U\left(\mathbf{e}_{s}\right)-U\left(\mathbf{e}_{2}\right)=0.73379
\end{gathered}
\end{equation}

In case 3, parameter values are $a=0.9,$ $b_{1}=1,$ $b_{2}=1,$ $c=2.5,$ $d_{1}=1.2,$ $d_{2}=1.2.$ $\mathbf{e}_{2}$ is more stable by metric 3 (depth) and metric 1 (linear stability), but not metric 2 (basin width) (Figure S5).
\begin{equation}
\begin{gathered}
\text{Metric 1: }\text{Re}\left(\lambda\left(\mathbf{e}_{1}\right)\right)=-1.79998,\quad
\text{Re}\left(\lambda\left(\mathbf{e}_{2}\right)\right)=-2.39984\\
\text{Metric 2: }\|\mathbf{e}_{1}-\mathbf{e}_{s}\|=1.81859,\quad
\|\mathbf{e}_{2}-\mathbf{e}_{s}\|=1.71693\\
\text{Metric 3: }U\left(\mathbf{e}_{s}\right)-U\left(\mathbf{e}_{1}\right)=0.837959,\quad
U\left(\mathbf{e}_{s}\right)-U\left(\mathbf{e}_{2}\right)=0.937962
\end{gathered}
\end{equation}

In case 4, the parameter values are $a=0.9,$ $b_{1}=0.9,$ $b_{2}=0.9,$ $c=2.37,$ $d_{1}=0.78,$ $d_{2}=1.46.$  $\mathbf{e}_{2}$ is more stable by metric 3 (depth), but not metric 1 (linear stability) or metric 2 (basin width) (Figure S6).
\begin{equation}
\begin{gathered}
\text{Metric 1: }\text{Re}\left(\lambda\left(\mathbf{e}_{1}\right)\right)=-1.61995,\quad
\text{Re}\left(\lambda\left(\mathbf{e}_{2}\right)\right)=-1.55978\\
\text{Metric 2: }\|\mathbf{e}_{1}-\mathbf{e}_{s}\|=1.80573,\quad
\|\mathbf{e}_{2}-\mathbf{e}_{s}\|=1.77222\\
\text{Metric 3: }U\left(\mathbf{e}_{s}\right)-U\left(\mathbf{e}_{1}\right)=0.81102,\quad
U\left(\mathbf{e}_{s}\right)-U\left(\mathbf{e}_{2}\right)=0.91103
\end{gathered}
\end{equation}

These four cases show that an equilibrium in a bistable system can be classified as ``more stable" by any combination of the three metrics. Hence it is important to recognize that each metric conveys a different piece of information about stability.
\newpage
\beginsupplement

\begin{figure}[ht]
\centering
\begin{subfigure}[b]{0.4\textwidth}
\caption{$\sigma=0.02$}
\includegraphics[width=\textwidth]{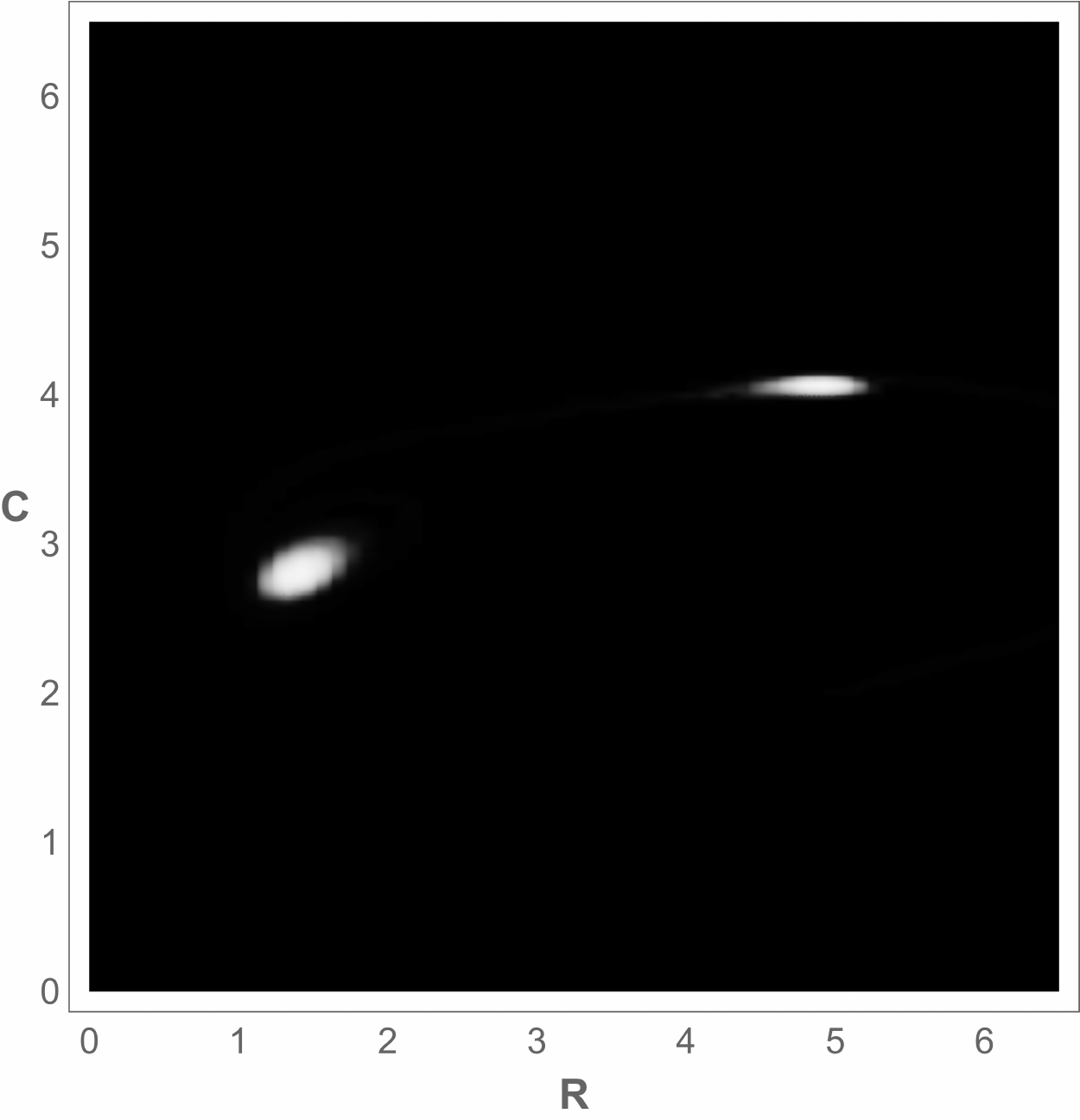}
\end{subfigure}
\begin{subfigure}[b]{0.4\textwidth}
\caption{$\sigma=0.1$}
\includegraphics[width=\textwidth]{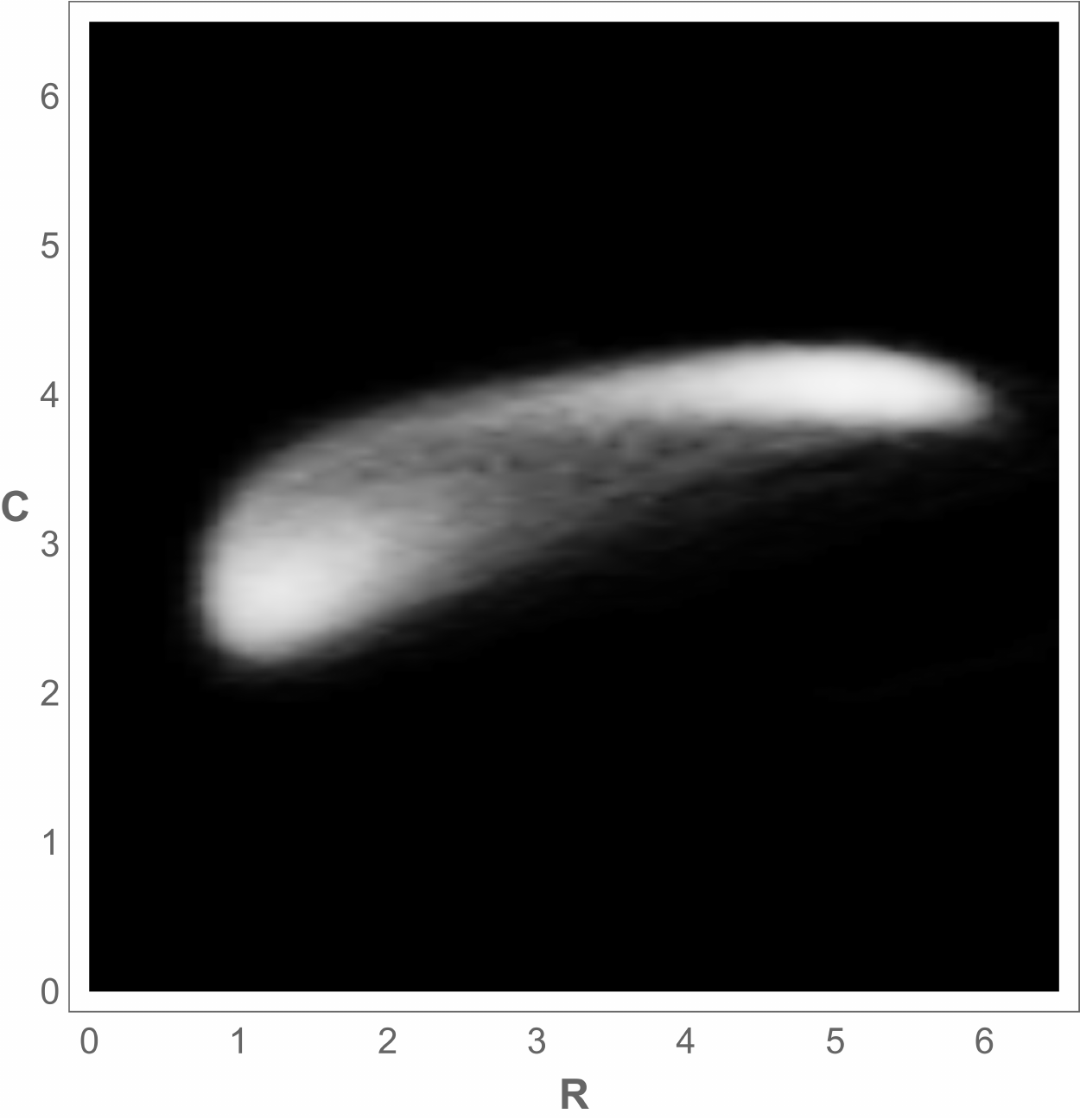}
\end{subfigure}
\vspace*{3mm}
\begin{subfigure}[b]{0.4\textwidth}
\caption{$\sigma=0.25$}
\includegraphics[width=\textwidth]{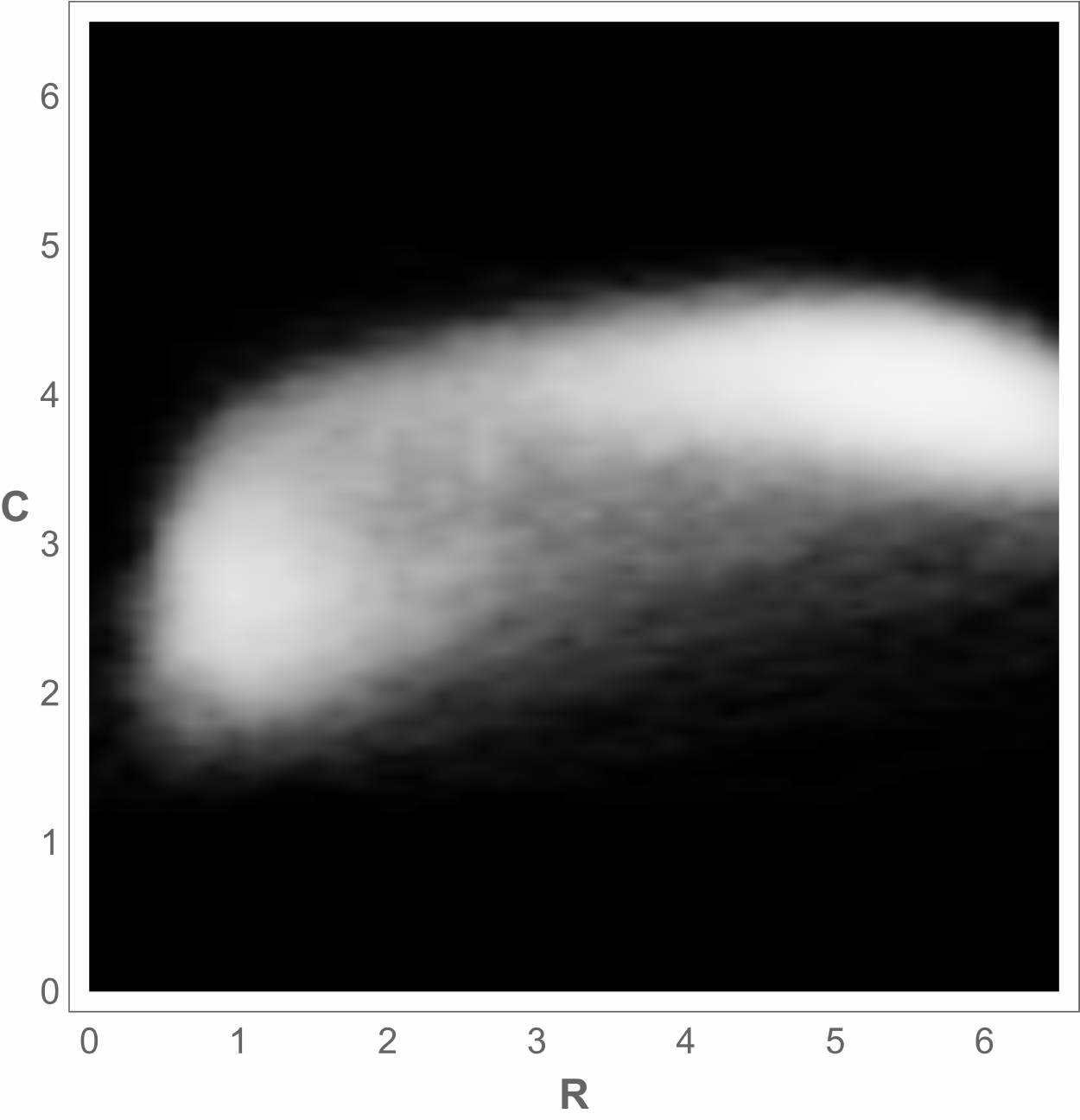}
\end{subfigure}
\begin{subfigure}[b]{0.4\textwidth}
\caption{$\sigma=0.5$}
\includegraphics[width=\textwidth]{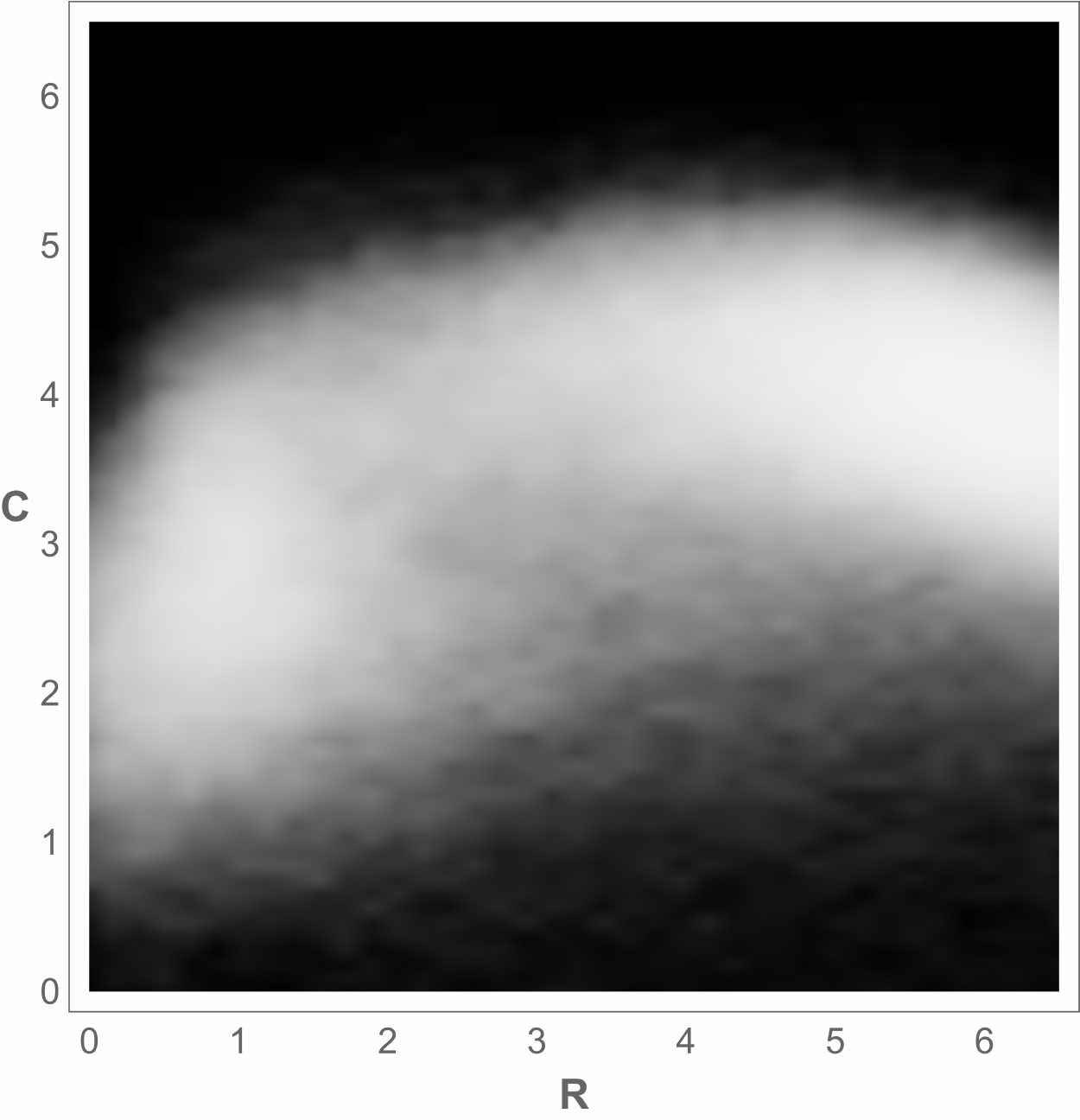}
\end{subfigure}
\caption{\small {\label{Fig3} Approximations to the steady-state probability density of equations \eqref{conres}, obtained from a long-time ($t=25000$) simulation, with four different noise intensities. Integration was performed with the Euler-Maruyama method and $\Delta t=0.025$. Variables are scaled, so the units are dimensionless. The horizontal axis is resource population density and the vertical axis is consumer population density. White corresponds to high probability density. The information conveyed in each plot depends on the noise intensity (see appendix~\ref{subsec:SmallNAppendix}).}}
\end{figure}

\begin{figure}[ht]
\centering
\includegraphics[width=\textwidth]{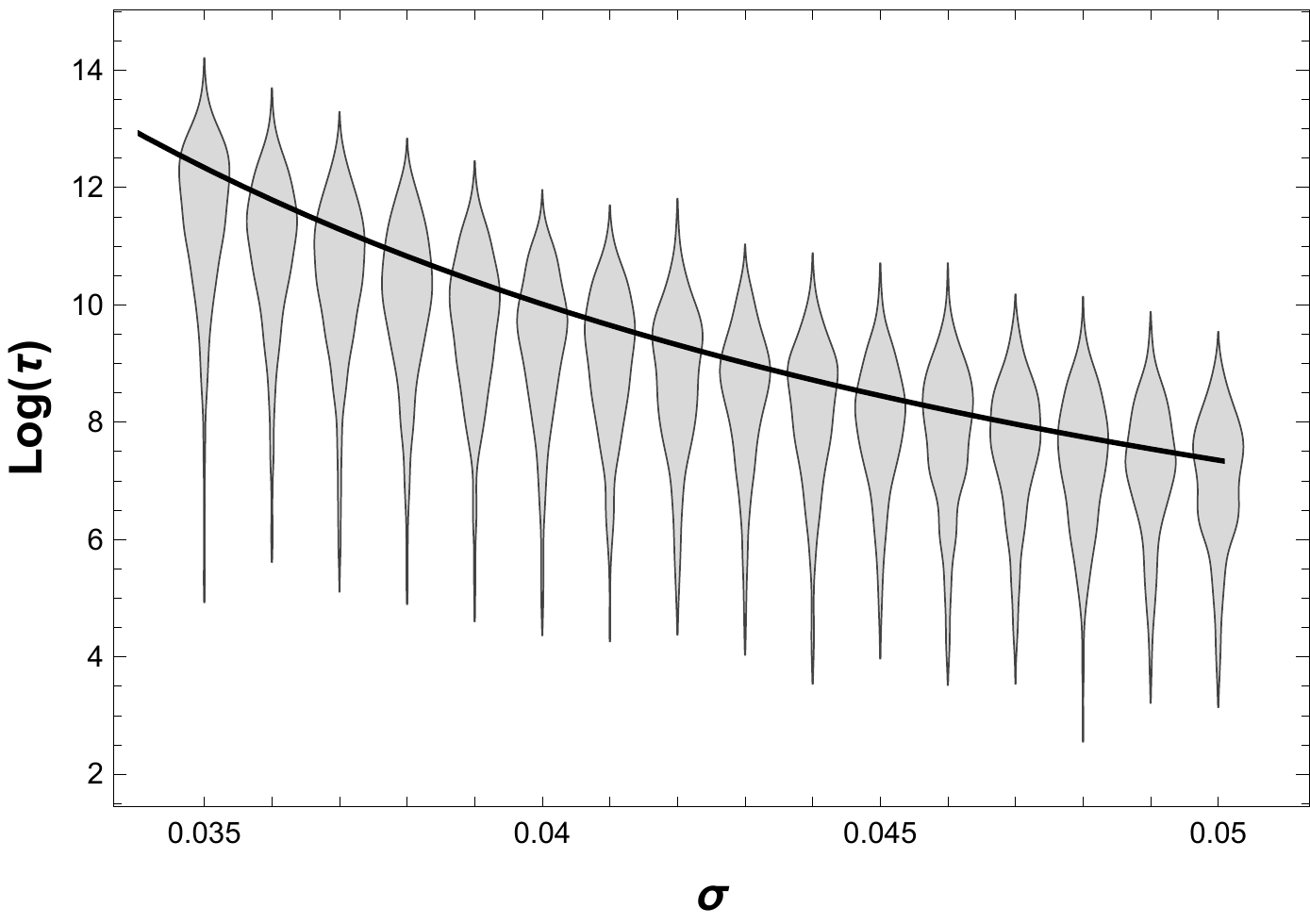}
\caption{\small {\label{Fig30} Simulation results for first passage times in example 2. The initial point is $\mathbf{e_{A}}$, and the time step is $0.05$. $500$ realizations were generated at each noise level. The width of each gray shape corresponds to the frequency with which each first passage time was observed. The black line is the small-noise approximation of the mean first passage time from the formula in appendix~\ref{subsec:MFPTa}. Note that the small-noise approximation matches the means of the distributions well. At all noise levels, the simulations included outliers that escaped from the basin of attraction much faster than the small-noise prediction.}}
\end{figure}

\begin{figure}[ht]
\centering
\begin{subfigure}[b]{0.7\textwidth}
\caption{}
\includegraphics[width=\textwidth]{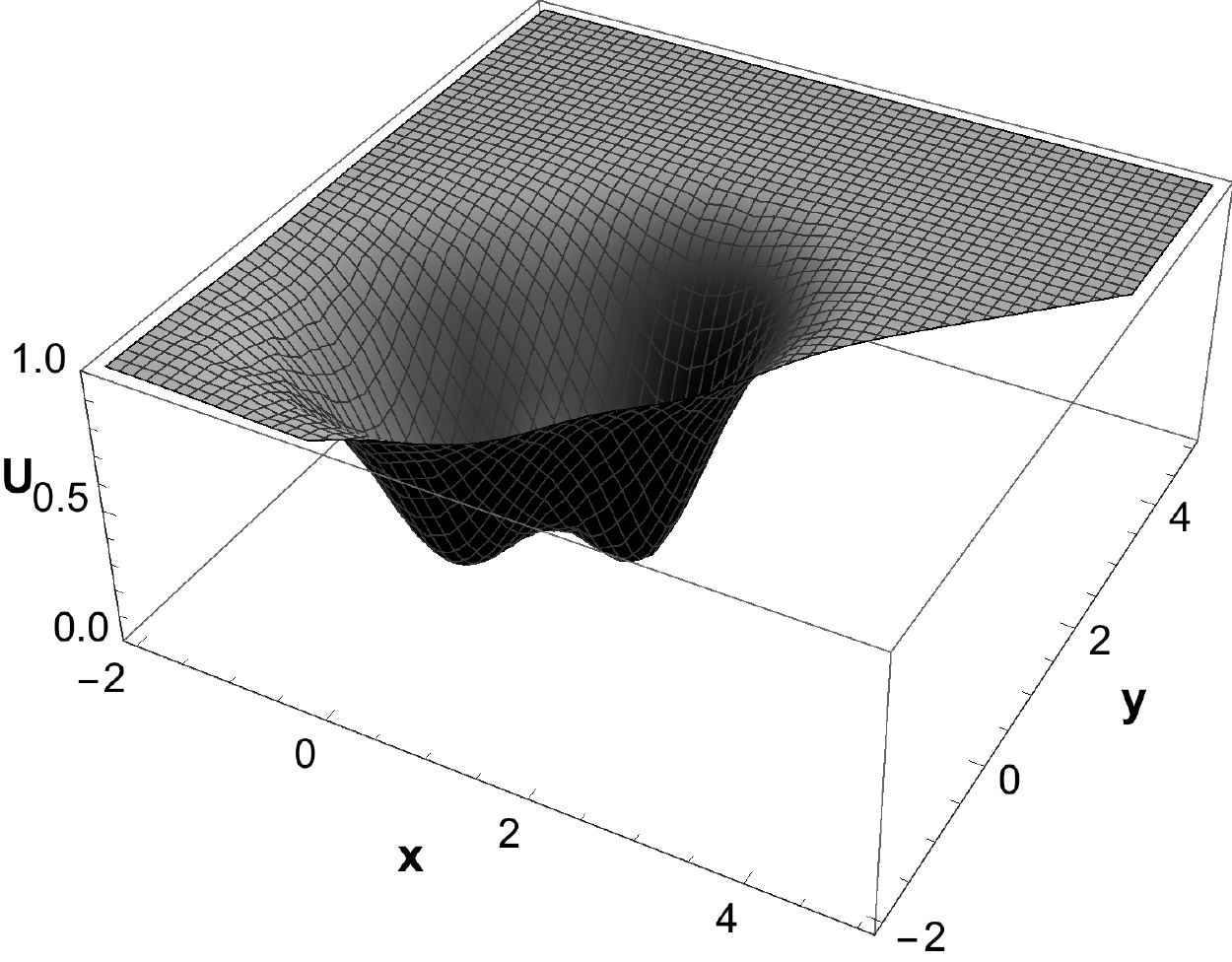}
\end{subfigure}
\vspace*{5mm}
\begin{subfigure}[b]{1.0\textwidth}
\caption{}
\includegraphics[width=\textwidth]{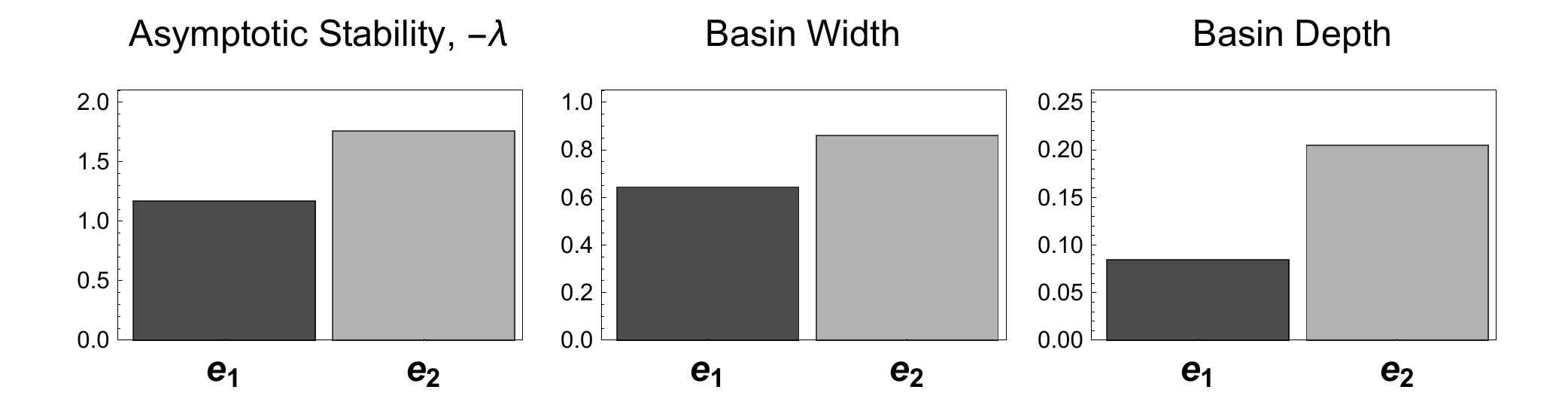}
\end{subfigure}
\caption{\small {\label{case1fig} \textbf{(a)} The potential function for case 1 of the system in appendix~\ref{subsec:AnotherEx}. \textbf{(b)} A comparison of three different metrics of stability for the system in case 1.}}
\end{figure}

\begin{figure}[ht]
\centering
\begin{subfigure}[b]{0.7\textwidth}
\caption{}
\vspace*{5mm}
\includegraphics[width=\textwidth]{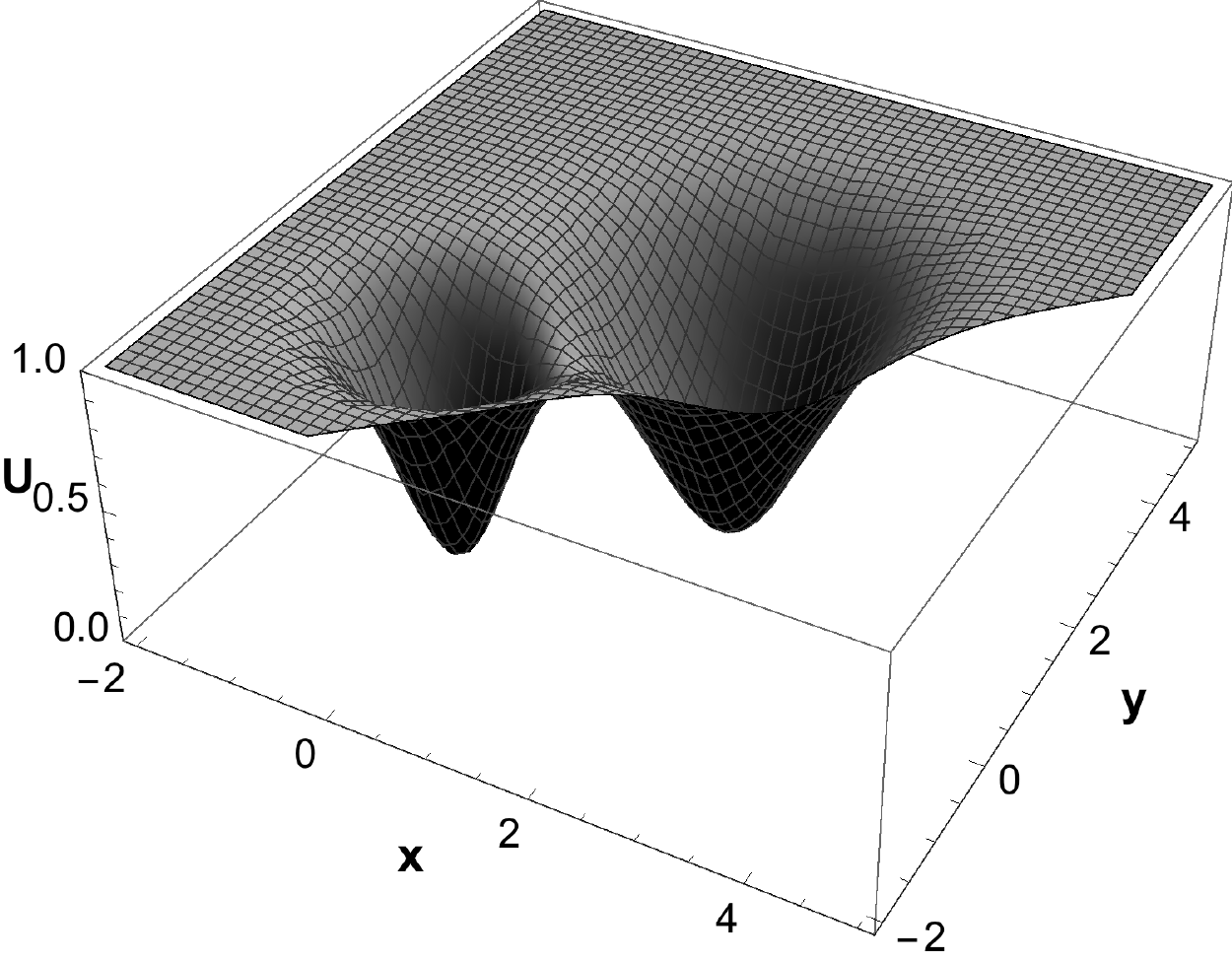}
\end{subfigure}
\begin{subfigure}[b]{1.0\textwidth}
\caption{}
\includegraphics[width=\textwidth]{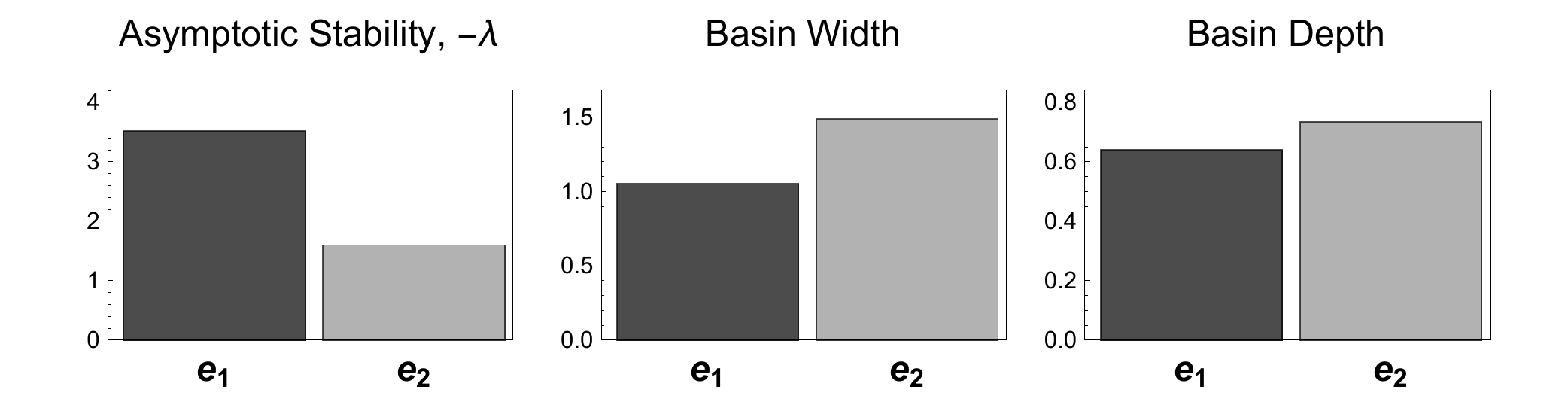}
\end{subfigure}
\caption{\small {\label{case2fig} \textbf{(a)} The potential function for case 2 of the system in appendix~\ref{subsec:AnotherEx}. \textbf{(b)} A comparison of three different metrics of stability for the system in case 2.}}
\end{figure}

\begin{figure}[ht]
\centering
\begin{subfigure}[b]{0.7\textwidth}
\caption{}
\vspace*{5mm}
\includegraphics[width=\textwidth]{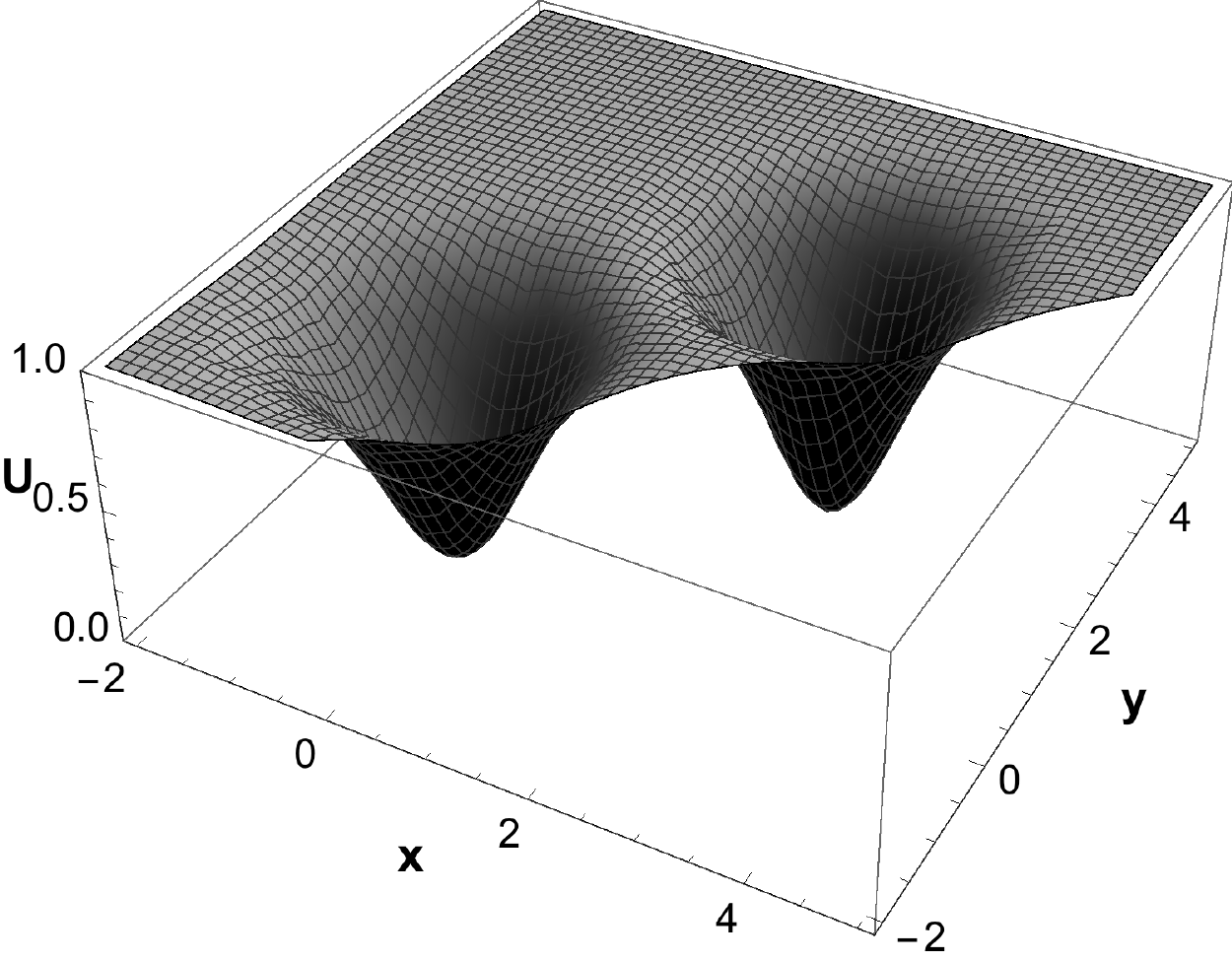}
\end{subfigure}
\begin{subfigure}[b]{1.0\textwidth}
\caption{}
\includegraphics[width=\textwidth]{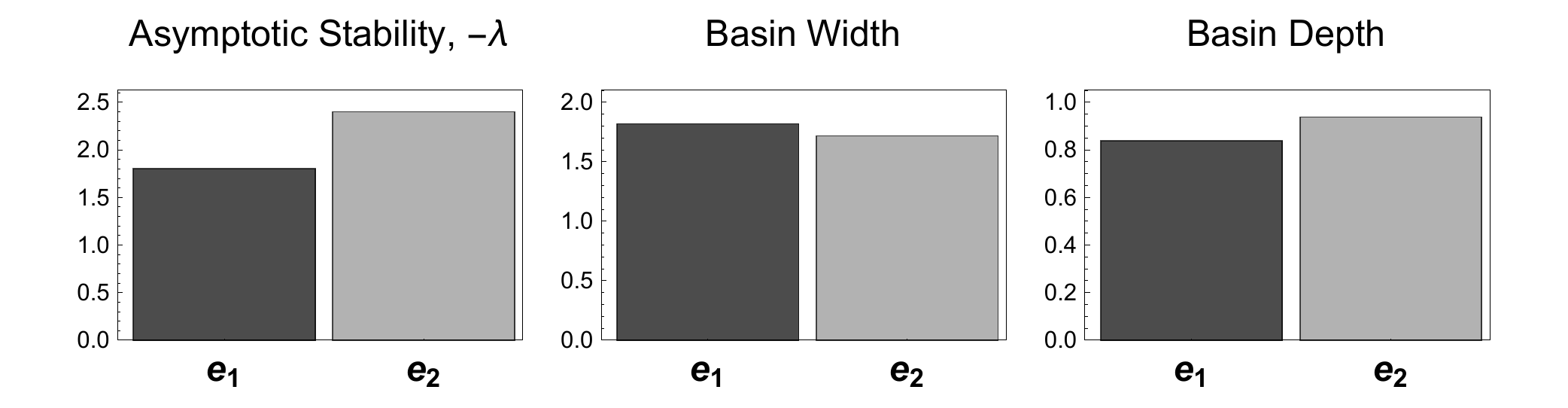}
\end{subfigure}
\caption{\small {\label{case3fig} \textbf{(a)} The potential function for case 3 of the system in appendix~\ref{subsec:AnotherEx}. \textbf{(b)} A comparison of three different metrics of stability for the system in case 3.}}
\end{figure}

\begin{figure}[ht]
\centering
\begin{subfigure}[b]{0.7\textwidth}
\caption{}
\vspace*{5mm}
\includegraphics[width=\textwidth]{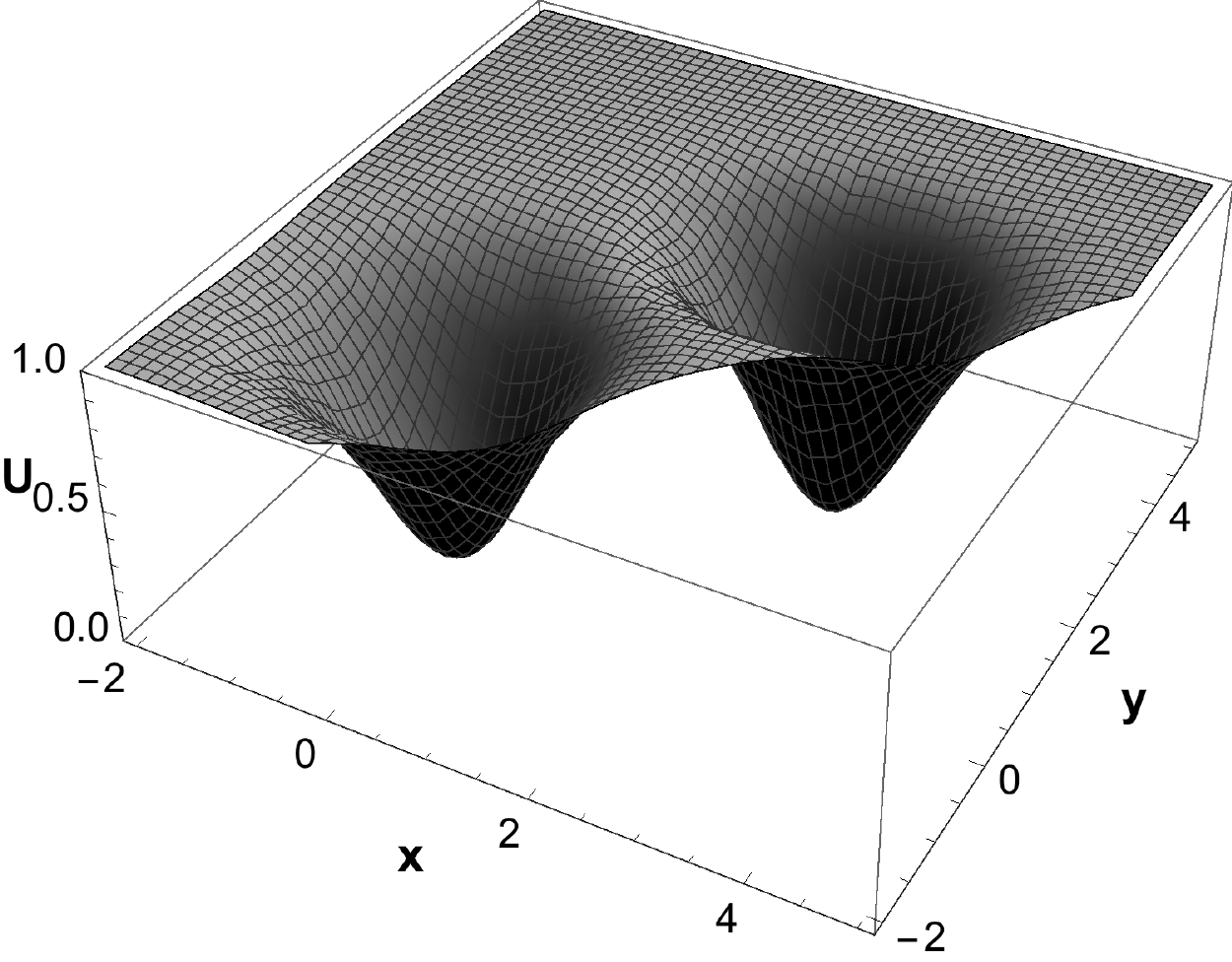}
\end{subfigure}
\begin{subfigure}[b]{1.0\textwidth}
\caption{}
\includegraphics[width=\textwidth]{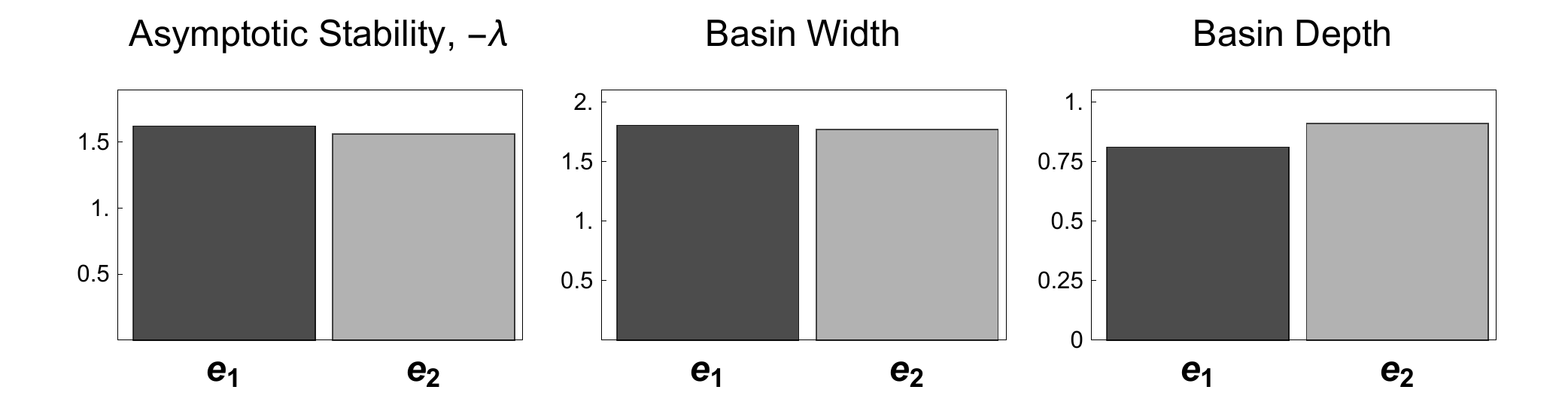}
\end{subfigure}
\caption{\small {\label{case4fig} \textbf{(a)} The potential function for case 4 of the system in appendix~\ref{subsec:AnotherEx}. \textbf{(b)} A comparison of three different metrics of stability for the system in case 4.}}
\end{figure}

\end{document}